\documentclass[sigconf, author]{acmart}
\usepackage{amsmath,amsfonts,dsfont,multirow,multicol,epsfig,url,array,makecell,balance,color,epstopdf}
\usepackage[ruled, vlined, linesnumbered]{algorithm2e}
\usepackage{hhline}
\usepackage{array}
\usepackage{enumerate}
\usepackage{enumitem}
\usepackage{subfigure}
\usepackage{booktabs}
\usepackage{cellspace}
\usepackage{xcolor,colortbl}
\usepackage{bm}
\usepackage{lipsum}
\usepackage{tabularx}
\usepackage{diagbox}
\usepackage{caption}
\usepackage{threeparttable}
\setlist[itemize]{leftmargin=*}
\PassOptionsToPackage{numbers, compress}{natbib}

\newtheorem{definition}{Definition}
\newtheorem{lemma}{Lemma}
\newtheorem{theorem}{Theorem}

\def\header{\vspace{0.8mm} \noindent}

\def\tblcapup{\vspace{0mm}}

\newcommand{\pushright}[1]{\ifmeasuring@#1\else\omit\hfill$\displaystyle#1$\fi\ignorespaces}
\newcommand{\pushleft}[1]{\ifmeasuring@#1\else\omit$\displaystyle#1$\hfill\fi\ignorespaces}

\newcommand{\eqn}[1]{{Equation~(#1)}}

\def\E{\mathrm{E}}

\def\odss{{ODSS}\xspace}
\def\dss{{Optimal Dynamic Subset Sampling}\xspace}
\def\bdss{{Basic Dynamic Subset Sampling}\xspace}

\def\naive{{the Naive method}\xspace}

\def\setsampler{{SetSampler}\xspace}

\def\G{\mathcal{G}}
\def\V{\mathcal{V}}
\def\E{\mathcal{E}}
\def\M{\mathcal{M}}
\def\Ex{\mathbb{E}}
\AtBeginDocument{%
  }

\copyrightyear{2023}
\acmYear{2023}
\setcopyright{acmlicensed}
\acmConference[KDD '23]{Proceedings of the 29th ACM SIGKDD Conference on Knowledge Discovery and Data Mining}{August 6--10, 2023}{Long Beach, CA, USA}
\acmBooktitle{Proceedings of the 29th ACM SIGKDD Conference on Knowledge Discovery and Data Mining (KDD '23), August 6--10, 2023, Long Beach, CA, USA}
\acmPrice{15.00}
\acmDOI{10.1145/3580305.3599458} 
\acmISBN{979-8-4007-0103-0/23/08}

\begin{document}

\title{Optimal Dynamic Subset Sampling: Theory and Applications}
\subtitle{[Technical Report]}

\author{Lu Yi}
\email{yilu@ruc.edu.cn}
\orcid{1234-5678-9012}
\affiliation{%
  \institution{Renmin University of China}
  \city{Beijing}
  \country{China}
  \postcode{100872}
}

\author{Hanzhi Wang}
\email{hanzhi_wang@ruc.edu.cn}
\affiliation{%
    \institution{Renmin University of China}
  \city{Beijing}
  \country{China}
  \postcode{100872}}

\author{Zhewei Wei}
\email{zhewei@ruc.edu.cn}
\authornote{Zhewei Wei is the corresponding author. The work was partially done at Gaoling School of Artificial Intelligence, Peng Cheng Laboratory, Beijing Key Laboratory of Big Data Management and Analysis Methods and MOE Key Lab of Data Engineering and Knowledge Engineering.}
\affiliation{%
    \institution{Renmin University of China}
  \city{Beijing}
  \country{China}
  \postcode{100872}}

 \begin{abstract}

We study the fundamental problem of sampling independent events, called {\em subset sampling}. Specifically, consider a set of $n$ distinct events $S=\{x_1, \ldots, x_n\}$, in which each event $x_i$ has an associated probability $p(x_i)$. The subset sampling problem aims to sample a subset $T \subseteq S$, such that every $x_i$ is independently included in $T$ with probability $p(x_i)$. A naive solution is to flip a coin for each event, which takes $O(n)$ time. However, an ideal solution is a data structure that allows drawing a subset sample in time proportional to the expected output size $\mu=\sum_{i=1}^n p(x_i)$, which can be significantly smaller than $n$ in many applications. The subset sampling problem serves as an important building block in many tasks and has been the subject of various research for more than a decade. 

However, the majority of existing subset sampling methods are designed for a {\em static setting}, where the events in set $S$ or their associated probabilities remain unchanged over time. These algorithms incur either large query time or update time in a {\em dynamic setting} despite the ubiquitous time-evolving events with varying probabilities in real life. Therefore, it is a pressing need, but still, an open problem, to design efficient dynamic subset sampling algorithms. 

In this paper, we propose \odss, the first optimal dynamic subset sampling algorithm. The expected query time and update time of \odss are both optimal, matching the lower bounds of the subset sampling problem. We present a nontrivial theoretical analysis to demonstrate the superiority of \odss. We also conduct comprehensive experiments to empirically evaluate the performance of \odss. Moreover, we apply \odss to a concrete application: Influence Maximization. We empirically show that our \odss can improve the complexities of existing Influence Maximization algorithms on large real-world evolving social networks.

\end{abstract}




\maketitle

\vspace{-2mm}
\section{Introduction} \label{sec:intro}

In the past decade, we have been experiencing a huge ``Big Data'' movement fueled by our ever-increasing ability and desire to gather, store, and share data. Driven by the exponential blowup in data volumes, efficient algorithms are now in higher demand than ever before. In particular, {\em sampling} is one of the most powerful techniques in algorithm design and analysis, which can effectively reduce the problem size and is often a necessary tool for achieving high scalability. On the other hand, randomly chosen samples also serve as robust estimators for critical quantities in many cases.



In this paper, we study the fundamental problem of sampling independent events, called {\em Subset Sampling}. Specifically, consider a set $S$ with $n$ elements $S=\{x_1, \ldots, x_n\}$. All of the $n$ elements represent $n$ distinct events, each of which (e.g., the $i$-th element $x_i$) is associated with a probability $p(x_i)\in [0,1]$. A query for the subset sampling problem returns a subset $T\subseteq S$, in which the $i$-th event $x_i$ is independently included in $T$ with probability $p(x_i)$. 
A trivial bound for the subset sampling problem is $O(n)$, which can be achieved by flipping a biased coin for every $p(x_i)$. However, such complexity can be significantly larger than the expected output size $\mu=\sum_{i=1}^n p(x_i)$, leading to excessive time cost. A well-adopted goal for the subset sampling problem is to draw a subset sample in time roughly proportional to the expected output size $\mu=\sum_{i=1}^n p(x_i)$. 

The subset sampling problem has long been a crucial building block in many tasks. Various applications are in dire need of efficient subset sampling techniques, especially of the implementations in {\em dynamic} settings. To be more specific, in a dynamic setting, the probability that an event happens can be updated dynamically, and the elements in set $S$ are allowed to be changed over time. Such dynamic settings are extremely common in real life, as events are usually time-evolving in practice. Nonetheless, designing dynamic subset sampling algorithms is a more challenging task due to its hardness. 
In the following, we will present three concrete examples to demonstrate the wide applications of the subset sampling problem and the dynamic settings of these applications. 

\begin{table} [t]
\centering
\renewcommand{\arraystretch}{1.4}
\tblcapup
\caption{Comparison of subset sampling algorithms.}\label{tbl:comparison}
\vspace{-2mm}
\resizebox{1\linewidth}{!}{
\begin{tabular}{ScScSc}
\toprule
{\bf Algorithm} &  {\bf Expected Query Time} & {\bf Update Time} \\ \midrule
The Naive Method & $O(n)$ & $O(1)$ \\
HybridSS~\cite{tsai2010heterogeneous} & $O\left(1+n\sqrt{\min\left\{\bar{p},1-\bar{p}\right\}}\right)$ & $O(n)$ \\
BringmannSS~\cite{bringmann2012efficient} & $O(1+\mu)$ & $O\left(\log^2{n}\right)$ \\
\odss (Ours) & $O(1+\mu)$ & $O(1)$ \\ \bottomrule
\end{tabular}
}
\vspace{-3mm}
\end{table}


\subsection{Concrete Applications}
\header{\bf Dynamic Influence Maximization. } 
The Influence Maximization (IM) problem aims to find a set of $k$ users in social networks which can infect the largest number of users in the network. At the heart of existing Influence Maximization algorithms is generating random reverse reachable (RR) sets efficiently. The state-of-the-art IM algorithm, SUBSIM~\cite{guo2020influence}, reduces the complexity of IM by leveraging a subset sampling approach~\cite{bringmann2012efficient} to accelerate the generation of RR sets. However, the subset sampling approach can only attain the optimality in a static setting, presenting a significant drawback due to its demanding $O(\log^2{n})$ update time when handling update operations. This complexity poses a substantial obstacle to the practical application of SUBSIM in real-world scenarios, where social influence is inherently dynamic. Take, for instance, the rapid surge in the influence of celebrities resulting from scandals or rumors within a short span of time. Such fluctuations can swiftly render the user influence rankings obsolete. Consequently, there exists an urgent imperative to devise an efficient subset sampling algorithm specifically tailored for dynamic settings. 


\header{\bf Approximate Graph Propagation. } In recent Graph Neural Network (GNN) research, an emerging trend is to employ node proximity queries to build scalable GNN models, including prominent examples such as SGC~\cite{wu2019simplifying}, APPNP~\cite{gasteiger2018predict}, PPRGo~\cite{bojchevski2020scaling}, and GBP~\cite{chen2010scalable}. These proximity-based GNNs, in contrast to the original GCN~\cite{kipf2016semi}, decouple prediction and propagation and thus enable the mini-batch training, leading to significant enhancements in model scalability. To model various proximity measures, Wang~\cite{wang2021approximate} proposes the unified graph propagation formulas and introduces a UNIFIED randomized algorithm to calculate the formulas efficiently. Notably, the algorithm frequently employs subset sampling techniques to enhance computation efficiency.

\header{\bf Computational Epidemiology. } Particle-based simulation models have become a prevalent choice in the field of computational epidemiology~\cite{germann2006mitigation}. In these models, each infector $I$ possesses the ability to independently infect susceptible individuals who have come into contact with $I$ during a specific time period $T$. Consequently, the infection process associated with each infector inherently presents itself as a subset sampling problem.

\subsection{Motivations and Contributions}
Despite the importance and wide-spread applications of the dynamic subset sampling problem, it remains an open problem to design an optimal dynamic subset sampling algorithm. 
To be more specific, it has been established that no subset sampling algorithm can run faster than $\Omega(1+\mu)$, where $\mu=\sum_{i=1}^n p(x_i)$ denotes the expected output size of the problem. 
And handling an update operation (e.g., inserting/deleting an element or modifying a probability) takes at least $\Omega(1)$ time. 
This implies that in the scenario of subset sampling, $\Omega(1+\mu)$ and $\Omega(1)$ are the lower bounds for query and update time complexity, respectively. 
Thus, an optimal dynamic subset sampling algorithm is required to achieve an $O(1+\mu)$ expected query time per sample with $O(1)$ update time per update operation. 
In other words, we look for a dynamic subset sampling algorithm that is able to derive the sampling result in time roughly proportional to the output size and only requires a constant number of operations to support real-time updates. 


\header{\bf Motivations. } The subset sampling problem has been the subject of extensive research for more than a decade~\cite{bringmann2012efficient,guo2020influence,tsai2010heterogeneous}. However, to the best of our knowledge, none of them can achieve optimality in a dynamic setting. More specifically, we summarize the complexities of existing subset sampling algorithms in Table~\ref{tbl:comparison}. 
Notably, the Naive method supports real-time updates, but its drawback lies in excessive query time, making it inefficient for handling large-scale datasets. Conversely, the BringmannSS method~\cite{bringmann2012efficient} builds dedicated index structures to reduce query time. Unfortunately, such index structures struggle to accommodate dynamic changes, leading to significant update time. 
As a consequence, the subset sampling problem remains open in a dynamic setting despite years of effort.

\header{\bf Contributions. } 
Motivated by the need to design efficient dynamic subset sampling algorithms, we make the following contributions. \begin{itemize}
\item We propose \odss, the first optimal dynamic subset sampling algorithm. We prove that our \odss only costs $O(1+\mu)$ expected query time and supports $O(1)$ update time per update operation. The two complexities are {\em both} optimal, matching the lower bounds of subset sampling. 

\item We conduct comprehensive experiments to empirically evaluate the performance of our \odss. The experimental results show that our \odss consistently outperforms all existing subset sampling algorithms on all datasets. 

\item To further demonstrate the effectiveness of our \odss, we apply our \odss to a concrete application: Influence Maximization. We empirically show that our \odss can improve the complexities of existing Influence Maximization algorithms on large real-world evolving social networks. 

\end{itemize}

\section{Preliminary} \label{sec:pre}

Considering the subset sampling problem, we note that the optimality can be trivially achieved when all probabilities in set $S$ are identical, i.e., $p(x_1)=p(x_2)=\cdots=p(x_n)=p$. This is because, in this special case, the index $j$ of the first sampled element follows the geometric distribution: $\Pr\left[j=i\right]=(1-p)^{i-1} \cdot p$. Given the memoryless property of geometric distribution, we can iterate the process of generating $i \sim {\rm{Geo}}(p)$ where ${\rm{Geo}}(p)$ is the geometric distribution with parameter $p$, and return $x_{j+i}$ as the second element sampled in set $S$. We repeat the above process until the index of the sampled element exceeds $n$. It is known that the random number $i \sim {\rm Geo} (p)$ can be generated in $O(1)$ time by setting $i=\left\lfloor\frac{\log \sf rand \rm()}{\log (1-{p})}\right\rfloor$, where $\sf rand \rm()$ denotes a uniform random number in $[0,1]$. Thus, in this special case, the expected query time is bounded by $O(1+np)=O(1+\mu)$ with $O(1)$ update time per update operation~\cite{knuth1981seminumerical,devroye2006nonuniform}. For reference, we call this method the GeoSS method.


In the general case, a trivial bound of query time complexity is $O(n)$ since we can flip a biased coin for every $p(x_i)$. We call such method the Naive method, which offers an $O(n)$ query time with an $O(1)$ update time per update operation. To further reduce the time cost, a series of studies have been devoted to this problem over the past decade. In the following, we will review the state-of-the-art algorithms and briefly analyze why these methods fall short of achieving optimality in a dynamic setting.

\header{\bf The HybridSS Method. } 
Tsai et al.~\cite{tsai2010heterogeneous} propose a subset sampling method called HybridSS, which achieves $O(1+n\sqrt{\min\{\bar{p}, 1-\bar{p}\}})$ expected query time with $O(n)$ update time in the worst case. Here $\bar{p}=\frac{1}{n}\sum_{i=1}^n p(x_i)$ denotes the mean of all probabilities. The idea of the HybridSS method is to divide the original set $S$ into two disjoint sets $X$ and $Y$, where $X=\{x_i\le \sqrt{\bar{p}} \mid x_i\in S\}$ and $Y=S - X$. The HybridSS method invokes the Naive method to sample elements in set $Y$. 
For the elements in set $X$, the HybridSS method first treats all probabilities in $X$ as $\sqrt{\bar{p}}$ and applies the GeoSS method in the special case to sample some candidates. Since $p(x_i)\le \sqrt{\bar{p}}$ for every $x_i \in X$, the HybridSS method then accepts each candidate (e.g., $x_i$) with probability $\frac{p(x_i)}{\sqrt{\bar{p}}}$. By this strategy, every element $x_i$ in set $X$ is still guaranteed to be included in the final subset $T$ with probability $\sqrt{\bar{p}} \cdot \frac{p(x_i)}{\sqrt{\bar{p}}}=p(x_i)$. 


\header{\bf The BringmannSS Method. } The BringmannSS method~\cite{bringmann2012efficient} is proposed by Bringmann and Panagiotou, which is the first algorithm achieving the optimal $O(1+\mu)$ expected query time complexity in a {\em static} setting. The core of the BringmannSS method is a bucket sort operation accompanied by the Alias method~\cite{walker1974new,walker1977efficient} for weighted sampling. Here weighted sampling is another crucial sampling schema that aims to sample an element from a probability distribution. The Alias method is the state-of-the-art weighted sampling approach (in a static setting, however), which achieves the optimal $O(1)$ time per sample by building a dedicated alias table. The BringmannSS method revealed an interesting interplay between the {\em subset sampling} problem and the {\em weighted sampling} problem. 

We describe the BringmannSS method in detail here since our \odss is partially inspired by the BringmannSS method. Specifically, the BringmannSS method first partitions all elements in set $S$ into $(\lceil \log{n} \rceil+1)$ buckets. The $k$-th bucket ($k=1, 2, \ldots, \lceil \log{n} \rceil$) consists of the element $x$ of which the associated probability $p(x)\in (2^{-k},2^{-k+1}]$. And the last bucket (i.e., the $(\lceil \log{n} \rceil+1)$-th bucket) consists of all the elements $x$ with $p(x)<\frac{1}{n}$. For each element $x$ in the $k$-th bucket, the BringmannSS method sets $\bar{p}(x)=2^{-k+1}$, which is actually an upper bound on $p(x)$. As a result, after partitioning elements into buckets, all elements $x\in S$ have been sorted in descending order of $\bar{p}(x)$. The BringmannSS method further considers the elements ranked in $[2^k, 2^{k+1})$ as a group, denoted as $B_k$. 
Then the BringmannSS method invokes the SOTA weighted sampling method, the Alias method, to find the groups in which at least one element is sampled. Specifically, starting at the first group $B_1$, the BringmannSS method invokes the Alias method to sample the group index $j$ from the probability distribution $\Pr\left[X=j\right]=q_j \cdot \Pi_{k=1}^{j-1} (1-q_k)$. Here $q_k$ denotes the probability that at least one potential element (sampled with probability $\bar{p}(x)$) is sampled in $B_k$ (i.e., $q_k=1-(1-\bar{p}(x))^{|B_k|}$). The BringmannSS method iterates the above process starting at $B_{j+1}$ to sample the next group until $j>\lceil \log{n} \rceil+1$. In each sampled group, the BringmannSS method adopts the idea given in the HybridSS method. That is, we can first sample a candidate $x$ with probability $\bar{p}(x)$, then accept $x$ with probability $\frac{p(x)}{\bar{p}(x)}$.

The BringmannSS method achieves $O(1+\mu)$ expected query time but incurs an $O(\log^2{n})$ update time per update operation. Despite the limitations, the BringmannSS method provides two key insights for reducing complexity: (i) We can partition the elements in set $S$ into groups and sample them at the group level first. By such a strategy, we can effectively reduce the problem size. (ii) There is a rich interplay between different sampling schemas. Some techniques adopted in other sampling problems may also contribute to the algorithm design for the subset sampling problem. 

\header{\bf Remark. }
After the acceptance of the conference version of this paper, we learned of an interesting concurrent algorithm \setsampler proposed by Bhattacharya et al. ~\cite{bhattacharya2023near}. We are pleased to find that the \setsampler algorithm is somewhat simpler than our \odss method, while the (amortized) query time and the update time of \setsampler are still optimal. Moreover, the \setsampler algorithm was further applied to the task of fractional (bipartite) matching in~\cite{bhattacharya2023near}, which also demonstrates that the dynamic subset sampling problem is an important primitive of both theoretical and practical interests. 

\subsection{Other Related Work} 
In this subsection, we briefly introduce a useful trick named \textit{table lookup}, which is widely adopted in algorithms for weighted sampling~\cite{hagerup1993maintaining, hagerup1993optimal, matias2003dynamic}. At the heart of the table lookup trick is a table structure. Each row index refers to a probability distribution, and cells in a row stores the sampling outcomes corresponding to the probability distribution. Consider a weighted sampling problem defined on a finite domain with bounded size of possible probability distributions. We can first use the given distribution to index into the table, and uniformly sample a cell in the corresponding row to derive the sampling outcome. Since all possible outcomes under all possible probability distributions have been included in the table, we can achieve the optimal $O(1)$ query time with $O(1)$ update time. 

\section{Algorithm}
{\label{sec:alg}}

We first propose \textit{\bdss} in Section~\ref{subsec:bdss}. In Section~\ref{sec:dss}, we improve this basic algorithm by the table lookup method and propose the optimal algorithm \textit{\dss (\odss)}, which achieves the optimal query time and the optimal update time in the meanwhile. 

\subsection{Basic Dynamic Subset Sampling}\label{subsec:bdss}

In the basic algorithm, the elements are divided into $O(\log n)$ groups, ensuring the probabilities of the elements within each group differ by a factor of at most $2$. Rather than querying within each group, we first perform queries at the group level and subsequently sample elements within the sampled groups. Note that the group-level querying itself constitutes a subset sampling problem with only $O(\log n)$ elements. By repeating this process iteratively, we continue to reduce the number of elements until it reaches a constant value, thereby obtaining the basic algorithm. The theoretical property of the basic algorithm will be analyzed in Section~\ref{sec:analysis}.

\begin{figure*}[t]
\begin{center}
\centerline{\includegraphics[width=\linewidth]{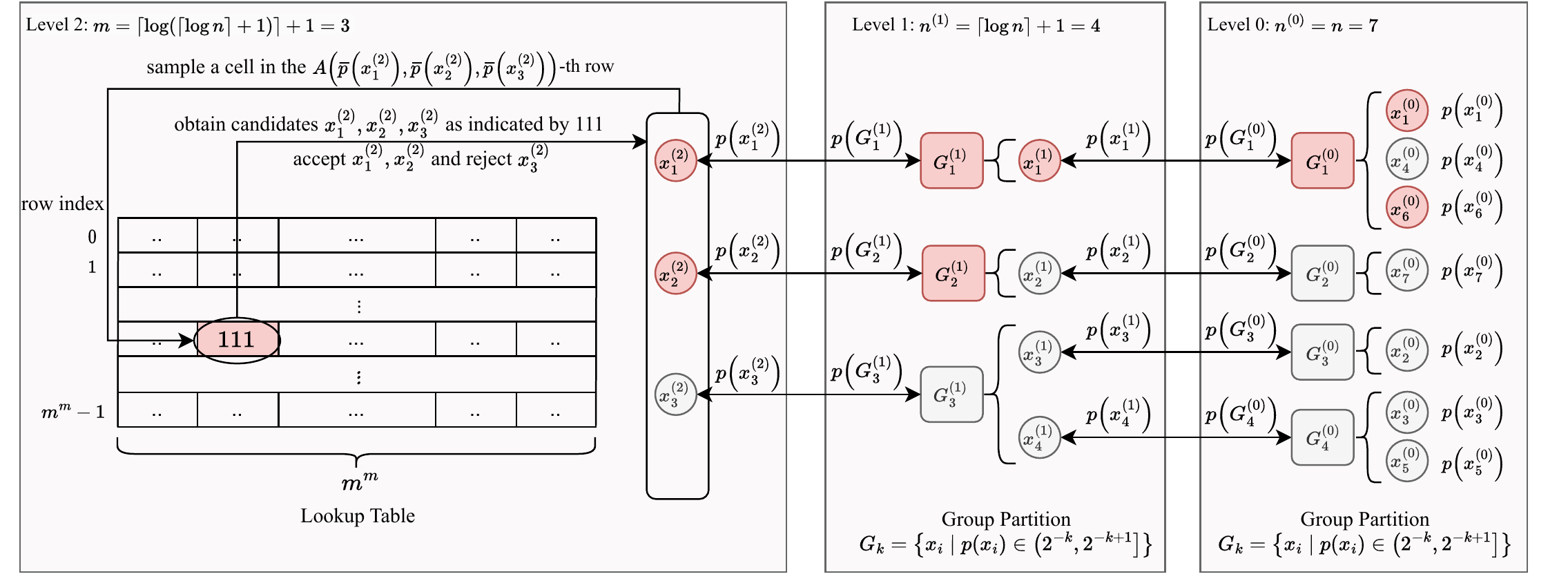}}
\vspace{-3mm}
\caption{An example of \dss.}
\label{fig:dss}
\vspace{-3mm}
\end{center}
\end{figure*}

\header{\bf Group Partitions. } We divide all the elements into $(\lceil \log n\rceil+1)$ groups. Define the $k$-th group $G_k=\lbrace x_i|2^{-k}<p(x_i)\le 2^{-k+1}\rbrace$ for $k\in \{1,\ldots, K-1\}$, and $G_K=\lbrace x_i|p(x_i)\le 2^{-K+1}\rbrace $, $K=\lceil \log n \rceil+1$. 
Note that $2^{-k+1}$ is the upper bound on $\{p(x_i)|x_i\in G_k\}$. 
Within the group $G_k$, we first sample elements as \textit{candidates} with probability $2^{-k+1}$ and then use rejection for each candidate to yield a correctly distributed sample.
Hence, the probability that there exists at least one candidate in $G_k$ can be calculated as $p(G_k)=1-(1-{2^{-k+1}})^{|G_k|}$. We first sample each group $G_k$ with probability $p(G_k)$ and then sample elements within each sampled group. 
Take the rightmost part of Figure~\ref{fig:dss} as an example. We divide seven elements into $K$ groups, where $K=\lceil \log n\rceil+1=4$. To distinguish with other symbols, we denote element $x_i$ as $x_i^{(0)}$ and group $G_i$ as $G_i^{(0)}$. $G_k^{(0)}$ contains elements with probabilities in $\left(2^{-k},2^{-k+1}\right]$ for $0\le k< K$. The last group $G_4^{(0)}$ contains elements with probabilities less than $2^{-3}$. After the division, $G_1^{(0)},G_2^{(0)},G_3^{(0)},G_4^{(0)}$ contain $\{x_1^{(0)},x_4^{(0)},x_6^{(0)}\},\{x_7^{(0)}\},\{x_2^{(0)}\},\{x_3^{(0)},x_5^{(0)}\}$, respectively.

\header{\bf Sampling within a Group. } 
We first consider the problem of sampling elements within a group $G_k$ given that $G_k$ is sampled.
Let $Y_k$ be an indicator random variable for the event that $G_k$ is sampled, that is, $\Pr\left[Y_k=1\right]=p(G_k)$. 
Note that the probability of each element must be converted to a conditional probability due to $Y_k=1$. 
We present Algorithm~\ref{alg:group} to illustrate the details of sampling element within a group $G_k$ conditioned on $Y_k=1$. 
Since we sample the elements as candidates with probability $2^{-k+1}$ first, the index $X_1$ of the first candidate in $G_k$ is geometrically distributed. 
That is, $\Pr[X_1=j]=2^{-k+1}(1-2^{-k+1})^{j-1}$. 
Then the conditional probability of the first candidate can be calculated as $$\Pr[X_1=j|Y_k=1]=\frac{\Pr[X_1=j\cap Y_k=1]}{\Pr[Y_k=1]}$$ for $j\in \{1,\ldots,|G_k|\}$.
Note that $\Pr[X_1=j\cap Y_k=1]$ is the same as $\Pr[X_1=j]$ since $X_1=j$ implies that $G_k$ contains at least one candidate. 
Thus, $\Pr[X_1=j|Y_k=1]=2^{-k+1}(1-2^{-k+1})^{j-1}/{p(G_k)}$. We generate a random number $r$, distributed as $X_1$ conditioned on $Y_k=1$, as the index of the first candidate of $G_k$ (Algorithm~\ref{alg:group} Line~\ref{line_first}). 
Subsequently, we proceed to sample the second candidate from the remaining group members. Note that the sampling of the second candidate is contingent upon two conditions: (1) $Y_k=1$ and (2) the $r$-th element is selected as the first candidate. Let $X_2$ be the index of the second candidate. Then we have $\Pr[X_2=j+r|Y_k=1 \cap X_1=r]=\Pr[X_2=j+r|X_1=r]=2^{-k+1}(1-2^{-k+1})^{j-1}$ since $X_1=r$ implies $Y_k=1$. Note that $X_2=j+r$ is the index in the whole group, while the index of the second candidate in the remainder of the group is $j$. 
Therefore, the index of the second candidate in the remainder of the group is distributed geometrically. 
We generate a new $r$, which is a geometric random variable, and select the $r$-th element in the remainder of the group as the second candidate (Algorithm~\ref{alg:group} Line~\ref{line_next}). 
Iterate the above process in $G_k$ to select the further candidates until the index exceeds $n_k=|G_k|$. After sampling the candidates, we accept each candidate $x_i$ with probability $p(x_i)/2^{-k+1}$. By this strategy, $x_i$ is still  guaranteed to be sampled with probability $2^{-k+1}\cdot \frac{p(x_i)}{2^{-k+1}}=p(x_i)$. The theoretical property of Algorithm~\ref{alg:group} will be shown in Lemma~\ref{lemma:group} in Section~\ref{sec:analysis}. Take the rightmost part of Figure~\ref{fig:dss} as an example again. Given that $G_1^{(0)}$ is sampled, we query in $G_1^{(0)}$ and obtain a sample $x_1^{(0)},x_6^{(0)}$. Since $G_2^{(0)},G_3^{(0)},G_4^{(0)}$ fail to be sampled, we do not query in these groups and thus no elements in these groups can be sampled.

\begin{algorithm}[t]
\caption{\sf SampleWithinGroup}
\label{alg:group}
\KwIn{a group $G_k$} 
\KwOut{a drawn sample $T$}
    $n_k \gets |G_k|$, $T \gets \emptyset$, $h\gets 0$\;
    Let $G_k[i]$ be the $i$-th element of $G_k$\;
    Generate a random $r$ s.t. $\Pr[r=j]=\frac{2^{-k+1}(1-2^{-k+1})^{j-1}}{p(G_k)}$, $j\in\{1,\ldots,n_k\}$\; \label{line_first}
    \While{$r+h\le n_k$}{
        $h\gets r+h$\;
        \If{${\sf rand}()<p(G_k[h])/2^{-k+1}$}{
            $T \gets T \cup \{G_k[h]\}$ \;
        }
        Generate a random $r\sim {\rm Geo}({2^{-k+1}})$\; \label{line_next}
    } 
\Return{ $ T $ }
\end{algorithm}

\header{\bf The Algorithm Structure. } Recall that querying at the group level is also a subset sampling problem with $O(\log n)$ elements. Each group $G_k$ is associated with probability $p(G_k)$. Partitioning the groups again can break down this subset sampling problem into a new subset sampling problem but with $O(\log \log n)$ elements. Continuing in this manner until the remaining subset sampling problem is of constant size and hence trivial to solve, we could derive the \textit{\bdss} algorithm as demonstrated in Algorithm~\ref{alg:bdss}.
To distinguish the subset sampling problems at various levels, we denote the set of elements at level $\ell$ as $S^{(\ell)}$ and the set of groups at level $\ell$ as $G^{(\ell)}$. Let $S^{(0)}=S=\{x_1,\ldots,x_n\}$. 
We first partition $S^{(0)}$ into a set of groups $G^{(0)}=\{G^{(0)}_1,\ldots, G^{(0)}_K\}$, $K=\lceil \log n\rceil+1$. 
The subset sampling problem for $G^{(0)}$ is defined at level $1$. 
Let $S^{(1)}=\{x_1^{(1)},\ldots,x_K^{(1)}\}$, $x_i^{(1)}$ with probability $p\big(G^{(0)}_i\big)$. 
Thus, a sampled $x_i^{(1)}$ indicates that $G_i^{(0)}$ is sampled. 
Continuing in this fashion, we maintain $\{S^{\ell}\}$ and $\{G^{\ell}\}$, $\ell\in \{0,\ldots,L\}$, until the size of $S^{(L)}$ is a constant, at which $L=\log^*n$.
For querying, we first sample each element $x^{(L)}_k\in S^{(L)}$ with probability $p\big(x^{(L)}_k\big)$ using the Naive method and enqueue the sampled elements into $Q^{(L)}$ (Algorithm~\ref{alg:bdss} Line~\ref{line:naive}). 
Note that each element at level $L$ corresponds to a group at level $L-1$, and the sampled element $x^{(L)}_k$ indicates that the group $G^{(L-1)}_k$ is sampled. 
Thus, for each sampled $x^{(L)}_k$ in $Q^{(L)}$, we sample the elements within $G^{(L-1)}_k$ by Algorithm~\ref{alg:group} and thus draw a sample of $S^{(L-1)}$. 
Iterating the above process from level $L-1$ down to level $1$, we finally obtain $Q^{(0)}$, a sample of $S^{(0)}$.

\begin{algorithm}[t]
\caption{\sf BasicDynamicSubsetSampling}
\label{alg:bdss}
\KwIn{the maximum level $L$, the set of elements $S^{(\ell)}$ and the set of groups $G^{(\ell)}$ at level $l$ for $0\le \ell \le L$} 
\KwOut{a drawn sample $T$}
$Q^{(\ell)}\gets \emptyset, 1\le\ell\le L$\;
Sample each element $x^{(L)}_k\in S^{(L)}$ with the Naive method and enqueue $x_k^{(L)}$ into $Q^{(L)}$ if $x_k^{(L)}$ is sampled\;\label{line:naive}
\For{ $\ell \gets L$ to 1 }{ \label{line:while}
    \While{ $ Q^{(\ell)} \neq \emptyset $ }{
        $x_k^{(\ell)} \gets$ {\sf deQueue}($Q^{(\ell)}$)\;
        Enqueue each element in the return of {\sf SampleWithinGroup}($G^{(\ell-1)}_k$) into $Q^{(\ell-1)}$\;\label{line:group}
    }
}
$T\gets Q^{(0)}$\;
\Return{ $T$ }
\end{algorithm}

\subsection{\dss}{\label{sec:dss}}
Although the basic algorithm described above demonstrates competitive query time, it falls short of meeting the lower bound. To ensure that the number of elements at level $L$ is a constant, we repeat the reduction $\log^*n$ times. It is remarkable that the number of groups becomes so small after a constant number of reduction steps. We present the table lookup method for solving the subset sampling problem with so few elements. Subsequently, we replace the Naive method with this table lookup method in the basic algorithm and reduce the reduction steps to $2$ times, thereby giving rise to the \textit{Optimal Dynamic Subset Sampling} algorithm (\odss).

\header{\bf Table Lookup. } 
We denote the total number of elements as $m$ to differentiate it from the original problem size, denoted as $n$. 
Consider a set $S$ with $m$ elements, $S=\{x_1,\ldots,x_{m}\}$. 
The subset sampling problem returns a subset of $S$ as a drawn sample, that is, each subset of $S$ is chosen as the drawn sample with a certain probability. 
We encode each subset of $S$ as a bit array $B$ with $m$ bits. 
$B[i]$ is the $i$-th (from right to left) bit indicating if $x_i$ is included in the subset. 
For instance, when $m=3$, $B= \sf \{011\}$ represents a subset $\{x_2,x_1\}$.
Since the elements are independently included in $B$, the probability of $B$ being the drawn sample is given by $p(B)=\prod_{x_i\in B}p(x_i)\cdot\prod_{x_j\notin B} (1-p(x_j))$. 
Notably, the sum of $p(B)$ over all distinct subsets $B\subseteq S$ is equal to $1$.
An important observation is that independently sampling each element $x_i$ with $p(x_i)$ and then returning the sampled elements is equivalent to returning a subset $B$ with probability $p(B)$. 
Note that the latter is a weighted sampling problem that is entirely distinct from subset sampling. The elements of this weighted sampling problem are subsets of $S$ and only one subset is returned as a sample.
Based on this observation, we apply the table lookup trick to solve the subset sampling problem for a special case, where $p(x_i)\in\{\frac{1}{m},\frac{2}{m},\ldots, \frac{m}{m}\}$. Note that $p(B)$ is a multiple of $\frac{1}{m^m}$ since $p(x_i)$ is a multiple of $\frac{1}{m}$, that is, $m^m p(B)\in \{1,\ldots,m^m\}$ for each subset $B$ of $S$. 
We create a \textit{lookup row} with $m^m$ entries and fill $m^mp(B)$ entries with $B$ for each subset $B$. The $m^m$ entries in the row are all filled with a subset since $p(B)$ for all subsets $B$ sum up to $1$ as mentioned above. 
When querying, we select an entry of the row uniformly and return the subset in the entry. There are $m^mp(B)$ entries filled with the subset $B$, so the probability of returning $B$ as a drawn sample is $m^mp(B)/m^m=p(B)$. 
For a general case without any limitation on $p(x_i)$, let $\bar{p}(x_i)=\lceil m p(x_i)\rceil/m$ for each element $x_i$. We maintain the lookup row with respect to $\bar{p}(x_i)$ instead of $p(x_i)$.
Thus, each entry in the row contains $x_i$ with probability $\bar{p}(x_i)$. Denote the elements in the subset as candidates. We accept each candidate with probability $p(x_i)/\bar{p}(x_i)$ to ensure that $x_i$ is sampled with $p(x_i)$.

\header{\bf Table Lookup for All Possible Distributions. }
If the probability of an element $x_i$ is altered, so too is $\bar{p}(x_i)$. 
To accommodate the modification, we also construct such a lookup row for each possible distribution of $(\bar{p}(x_1),\ldots,\bar{p}(x_{m}))$. 
Note that $\bar{p}(x_i)\in\{\frac{1}{m},\ldots, \frac{m}{m}\}$, so there are exactly $m^m$ different distributions of $ ( \bar{p}(x_1), \ldots, \bar{p}(x_{m}) ) $. 
To help with indexing into the table, let $$A(\bar{p}(x_1),\ldots,\bar{p}(x_{m}))=\sum_{i=1}^m (m\bar{p}(x_i)-1) m^{i-1}.$$
Note that $A ( \bar{p}(x_1), \ldots, \bar{p}(x_m) ) \in \{ 0, \ldots, m^m-1 \} $. 
For each distribution of $(\bar{p}(x_1), \ldots, \bar{p}(x_m))$, we fill in the $ A ( \bar{p}(x_1), \ldots, \bar{p}(x_m) ) $-th row with the sampling outcomes. 
When querying, we generate a random $r$ uniformly from $\{0,\ldots,m^m-1\}$, and return the subset in the entry in the $A(\bar{p}(x_1),\ldots,\bar{p}(x_m))$-th row and the $r$-th column.

\header{\bf The Optimal Algorithm Structure. } Replacing the Naive method in the basic algorithm with the table lookup trick, we finally derive the optimal algorithm, \dss (\odss). We provide the pseudocode in Algorithm~\ref{alg:dss}. The subset sampling problems at level $0$ and level $1$ are solved by querying within the sampled groups, while a table is used for querying at level $2$. When querying, we first sample $S^{(2)}$ using the table lookup method, which plays a role as the Naive method in the basic algorithm. The remains of \odss are the same as Lines~\ref{line:while} to ~\ref{line:group} in Algorithm~\ref{alg:bdss}, but with $L=2$. 

\begin{algorithm}[t]
\caption{\dss}
\label{alg:dss}
\KwIn{the set of elements $S^{(0)}, S^{(1)}, S^{(2)}$ and the set of groups $G^{(0)},G^{(1)}$, the table for $S^{(2)}$}
\KwOut{a drawn sample $T$}
$L\gets 2, T\gets \emptyset$\;
$Q^{(\ell)}\gets \emptyset, 1\le\ell\le L$\;
Draw a sample of $S^{(2)}$ by the table lookup method and enqueue the sampled elements into $Q^{(2)}$\;
The steps are the same as Algorithm~\ref{alg:bdss} Line ~\ref{line:while} to ~\ref{line:group} with $L=2$\;
$T\gets Q^{(0)}$\;
\Return{ $T$ }
\end{algorithm}

Figure~\ref{fig:dss} provides an example of \dss. There are seven elements $S=\{x_1,\ldots,x_7\}$. Let $S^{(0)}=S$. After the group partitions and maintaining the lookup table, we obtain the structure with three levels. At level $2$, we have $m=3$ and $S^{(2)}=\big\{x_1^{(2)},x_2^{(2)},x_3^{(2)}\big\}$. When querying, we first index into the $A \big( \bar{p}\big(x_1^{(2)}\big), \bar{p}\big(x_2^{(2)}\big), \bar{p}\big(x_3^{(2)}\big) \big) $-th row. By selecting an entry uniformly from $m^m$ entries in the row, we obtain a subset $B=\sf \{111\}$, which indicates $x_3^{(2)},x_2^{(2)},x_1^{(2)}$ are candidates. Then we accept each candidate $x_i^{(2)}$ with probability $p\big(x_i^{(2)}\big)\big/\bar{p}\big(x_i^{(2)}\big)$. It comes out that $x_1^{(2)},x_2^{(2)}$ are accepted, which implies that $G^{(1)}_1,G_2^{(1)}$ are sampled. Thus, we query within $G^{(1)}_1$ and $G_2^{(1)}$ using Algorithm~\ref{alg:group} and only $x_1^{(1)}$ in $G_1^{(1)}$ is sampled. Then, by querying within $G_1^{(0)}$, we accept $x_1^{(0)},x_6^{(0)}$. Therefore, we draw a sample of $S$, $\{x_1,x_6\}$. 

\subsection{Update Operations}\label{subsec:update}
In this subsection, we demonstrate how \odss handles element insertions, element deletions, and probability modifications. Consider the example of inserting an element $x$ with probability $p(x)$ into the set $S$. We start by adding $x$ to a group $G_k^{(0)}$ based on $p(x)$. Subsequently, we recalculate the probability $p\big(G_k^{(0)}\big)$ for this group.
The modification of $p\big(G_k^{(0)}\big)$ may necessitate transferring $x_k^{(1)}$ from one group at level $1$ to another, resulting in modifications of the probabilities of two groups at level $1$. In other words, the probabilities of two elements in $S^{(2)}$ are altered.
Denote the two elements at level $2$ as $x_i$ and $x_j$. Denote their new probabilities as $p'(x_i)$ and $p'(x_j)$,respectively. The row index at level $2$ can be revised by adding $(m\bar{p}'(x_i)-m\bar{p}(x_i))m^{i-1}+(m\bar{p}'(x_j)-m\bar{p}(x_j))m^{j-1}$ to the previous value $A(\bar{p}(x_0), \ldots, \bar{p}(x_m))$. This step completes the insertion of $x_i$. Therefore, each element insertion can be solved in constant time.



Considering an element deletion, we remove the element from the group it belongs to, which involves a modification of one group at level $0$. 
Then, we modify the probabilities of two elements at level $1$ and the row index at level $2$ with the same techniques as element insertions.
A probability modification can be achieved by an element insertion following an element deletion. 
Thus, all kinds of update operations can be solved in constant time. By maintaining the data structures correctly, we are able to achieve $O(n)$ memory cost and constant update time in the meanwhile. We defer the details of the data structures to Appendix~\ref{sec:app_structure}. 

Take Figure~\ref{fig:dss} as an example again. We update the set of elements by inserting $x_8$ to $S^{(0)}$ with $p(x_8)=0.3$. Then, $x_8$ is assigned to $G_2^{(0)}$ since $2^{-2} < p(x_8)\le 2^{-1}$. The size of $G^{(0)}_2$ is enlarged to two elements, so $p\big(G^{(0)}_2\big)$ is revised to $1-(1-2^{-1})^2=3/4$. Thus, $x_2^{(1)}$ has to be moved from $G^{(1)}_2$ to $G^{(1)}_1$. As a result, we have to revise $p\big(G_1^{(1)}\big), p\big(G_2^{(1)}\big)$. Subsequently, at level $2$, we recalculate $\bar{p}(x_1)$ and $\bar{p}(x_2)$, and revise $A(\bar{p}(x_1),\bar{p}(x_2),\bar{p}(x_3))$ accordingly. Then we are done with the insertion.

\section{Theoretical Analysis}{\label{sec:analysis}}
In this section, we analyze the theoretical properties of our algorithms. Theorem~\ref{theorem:bdss} and Theorem~\ref{theorem:dss} illustrate the correctness and the expected cost of \bdss and \dss, respectively. 

Our theoretical analysis is conducted under the standard word RAM model proposed by Fredman et al.~\cite{fredman1993surpassing}. The word RAM model is a commonly adopted computational model which offers a simple yet accurate abstraction of real-world computers. In particular, we will assume that any basic arithmetical operations on a single word of $\log n$ bits take constant time, where $n$ matches the problem size. The basic arithmetical operations include addition, multiplication, comparison, logical shifts, $\exp(x)$, and $\log(x)$ (the binary logarithm of $x$). The operation
\textit{\sf rand\rm() } (to generate a random number uniformly from $[0,1]$) takes constant time, too.

\begin{theorem}[\bdss]{\label{theorem:bdss}}
Using Algorithm~\ref{alg:bdss}, a subset sampling problem with $n$ elements can be solved in $O(2^{\log^*n }\cdot\mu+2^{\log^*n }+\log^*n)$ expected query time, $ O(n\log^*n) $ preprocessing time, and $ O (n\log^*n) $ space.
\end{theorem}

\begin{theorem}\label{theorem:dss}
Using Algorithm~\ref{alg:dss}, a subset sampling problem with $n$ elements can be solved in $O(1+\mu)$ query time, constant update time, $O(n)$ preprocessing time, and $O(n)$ space. 
\end{theorem}

To prove Theorem~\ref{theorem:bdss} and Theorem~\ref{theorem:dss}, we need the help of the following lemmas. For the sake of readability, we defer all proofs to Appendix~\ref{proof:group}~--~\ref{proof:dss}.
\begin{lemma}{\label{lemma:group}}
Given that a group $G_k=\lbrace x_i|2^{-k}<p(x_i)\le 2^{-k+1}\rbrace$ for $1\le k<K$ is successfully sampled with probability $p(G_k)=1-(1-2^{-k+1})^{n_k}$, Algorithm~\ref{alg:group} returns a sample $T$ of $G_k$ in $O(\mu_k+1)$ expected query time, where $K=\lceil \log n \rceil+1$, $\mu_k=\sum_{x_i\in G_k} p(x_i)$, $n_k=\left|G_k\right|\le n$. Each element $x_i\in G_k$ is included in $T$ independently with probability $p(x_i)$. A similar conclusion holds for $G_K=\{x_i|p(x_i)\le 2^{-K+1}\}$.
\end{lemma}

\begin{lemma}\label{lemma:mu}
Let $\mu=\sum_{i=1}^{n}p(x_i)$. Denote the sum of the probabilities for all elements in level $\ell$ as $\mu^{(\ell)}$. We have
\begin{equation}
\mu^{(\ell)}\le 2^\ell\mu+2^\ell-1. \label{eq:mu}
\end{equation}
\end{lemma}

\begin{lemma}\label{lemma:table}
With the table lookup method, the subset sampling problem with $S=\{x_1,\ldots,x_{m}\}$ can be solved with $O(1+\mu)$ query time, $O(2^m\cdot m+m^{2m})$ preprocessing time, and $O(m^{2m})$ space.
\end{lemma}

\section{Experiments} \label{sec:exp}

\begin{figure*}[t]
\begin{minipage}[t]{1\textwidth}
\centering
\begin{tabular}{cccc}
\hspace{-20mm} 
\includegraphics[width=44mm]{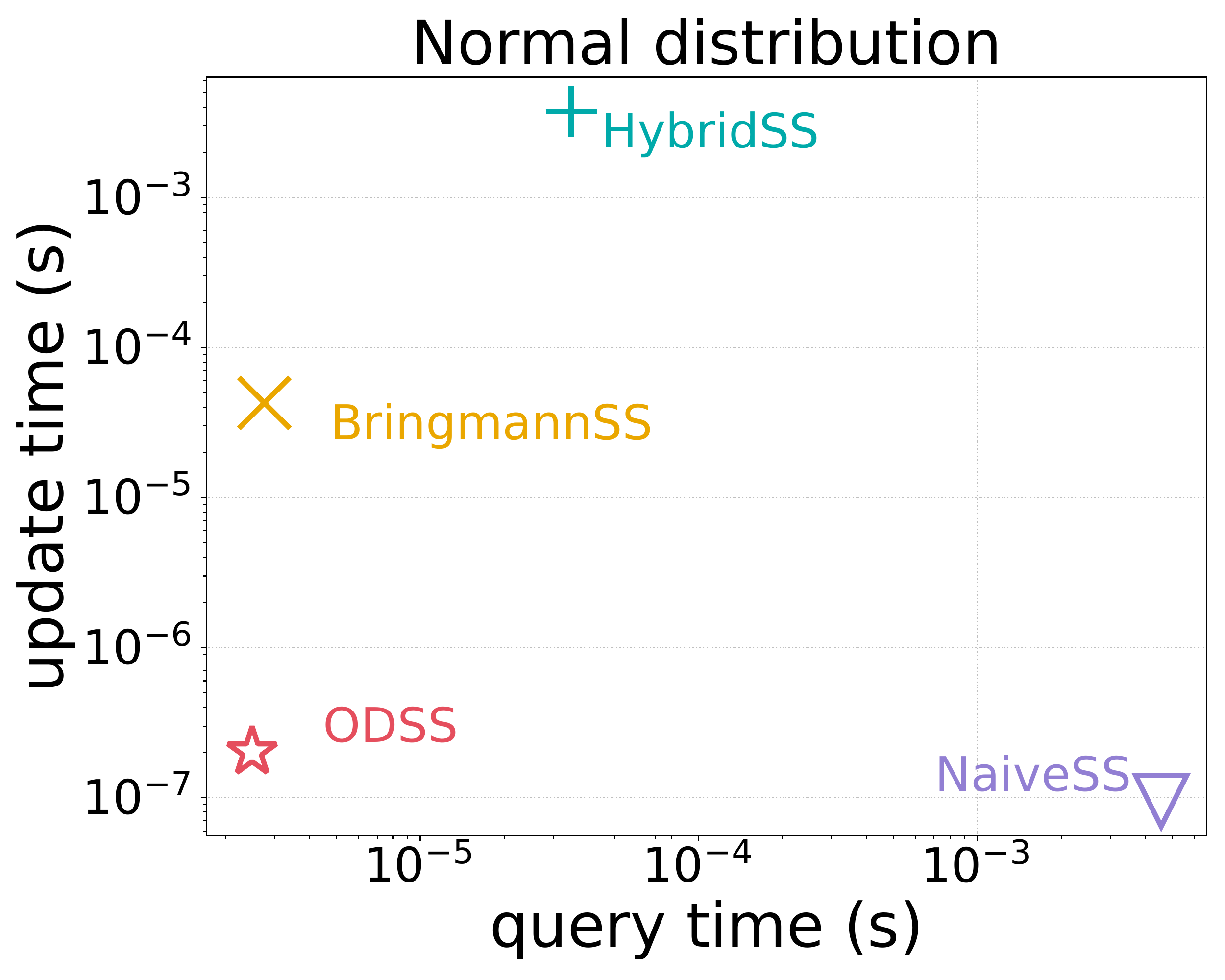}&
\hspace{-2mm} \includegraphics[width=44mm]{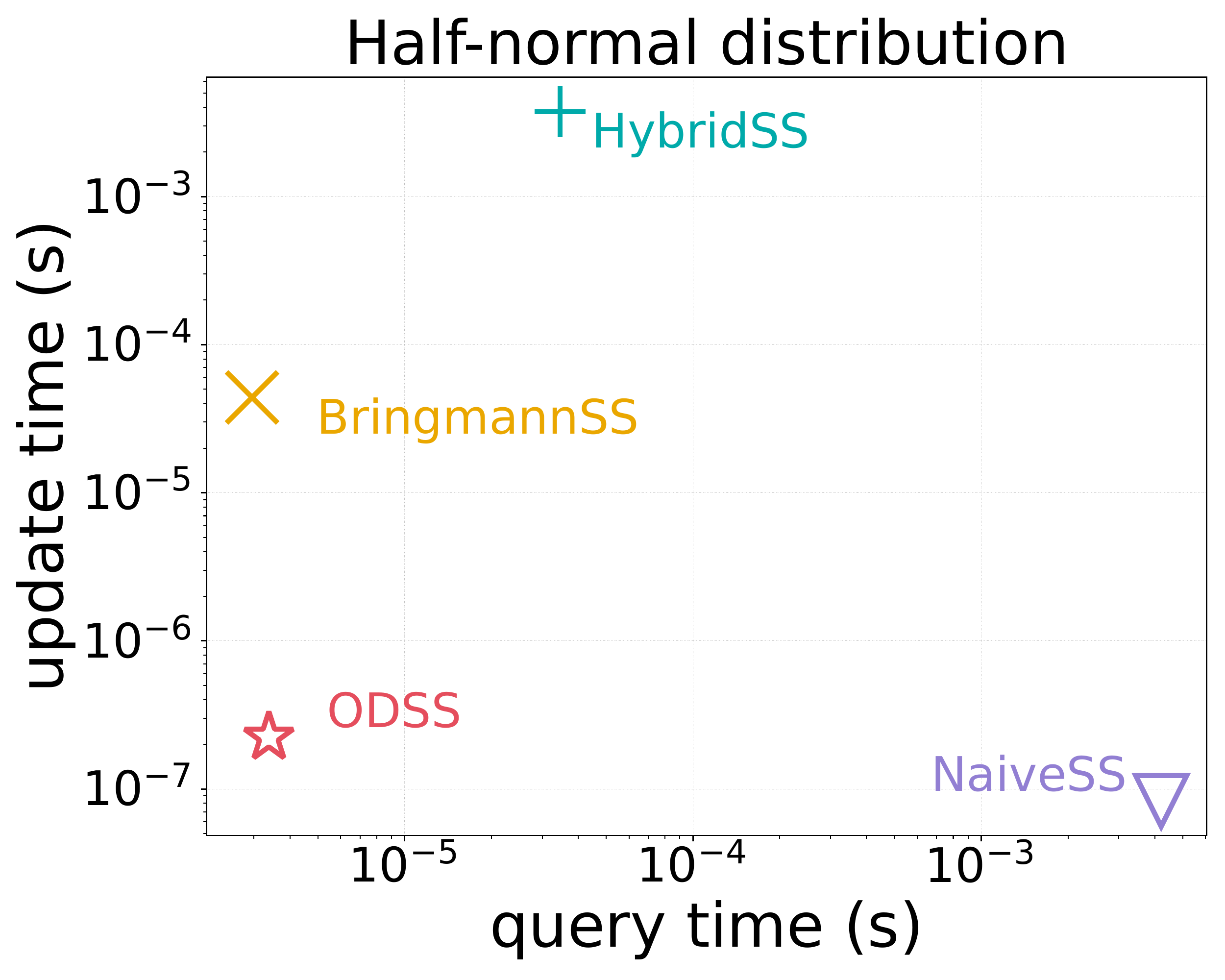} &
\hspace{-2mm} \includegraphics[width=44mm]{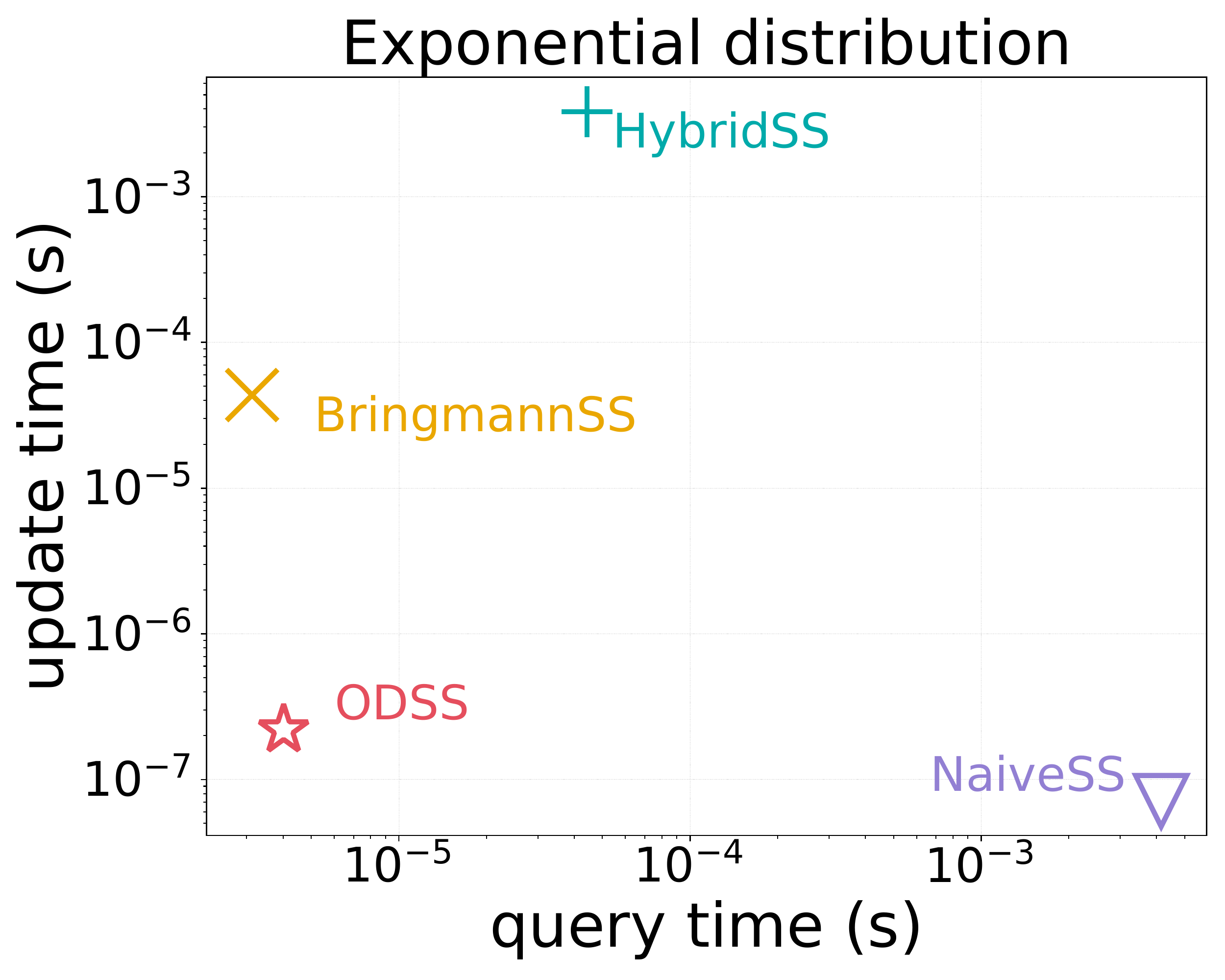} &
\hspace{-2mm} \includegraphics[width=44mm]{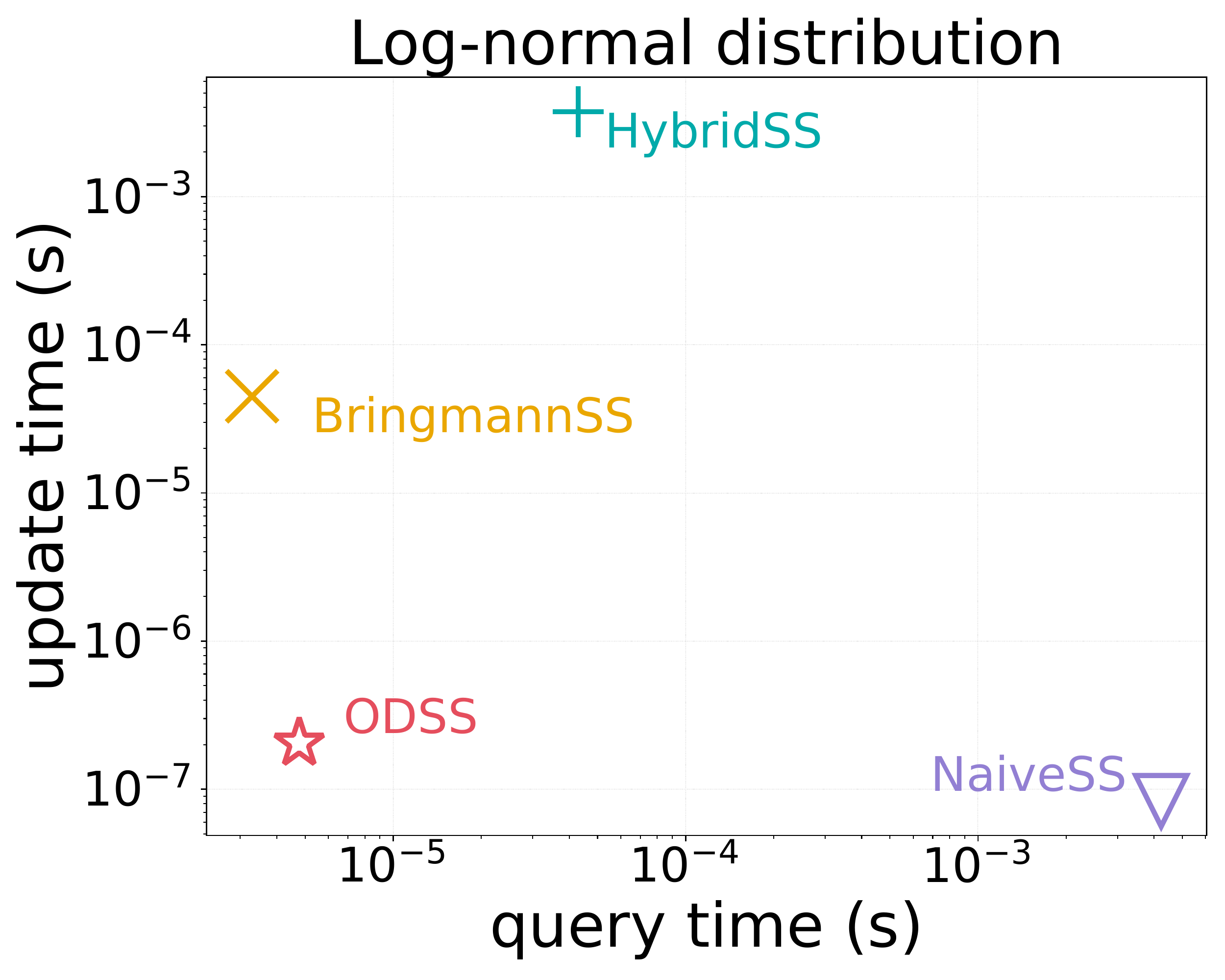}
\hspace{-20mm}
\end{tabular}
\vspace{-5mm}
\caption{Query time v.s. update time overhead on distributions with different skewnesses. ($n=10^5,\mu=1$)}
\label{fig:tradeoff}
\end{minipage}

\begin{minipage}[t]{1\textwidth}
\centering
\vspace{1mm}
\begin{tabular}{cccc}
\hspace{-20mm} 
\includegraphics[width=44mm]{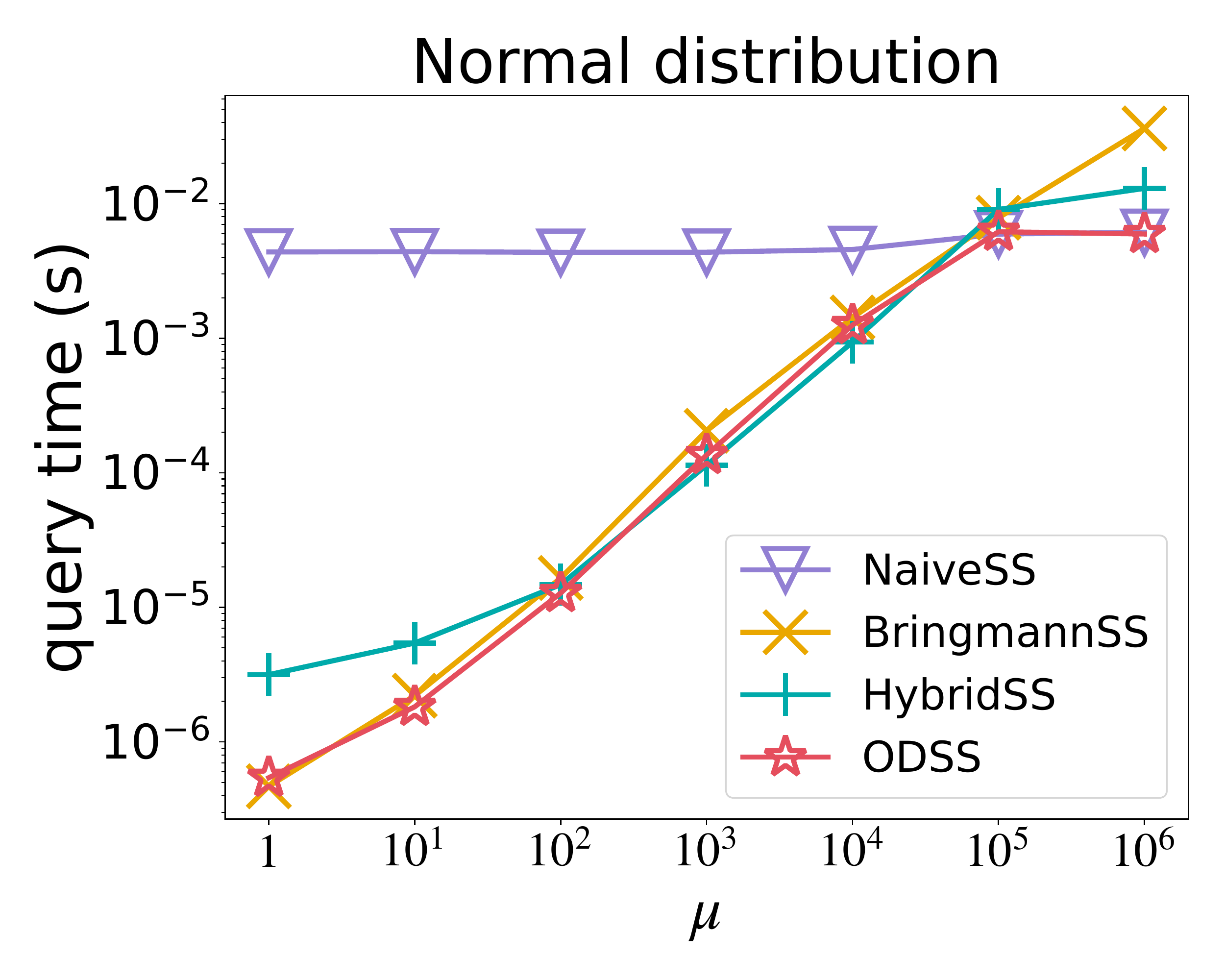}&
\hspace{-2mm} \includegraphics[width=44mm]{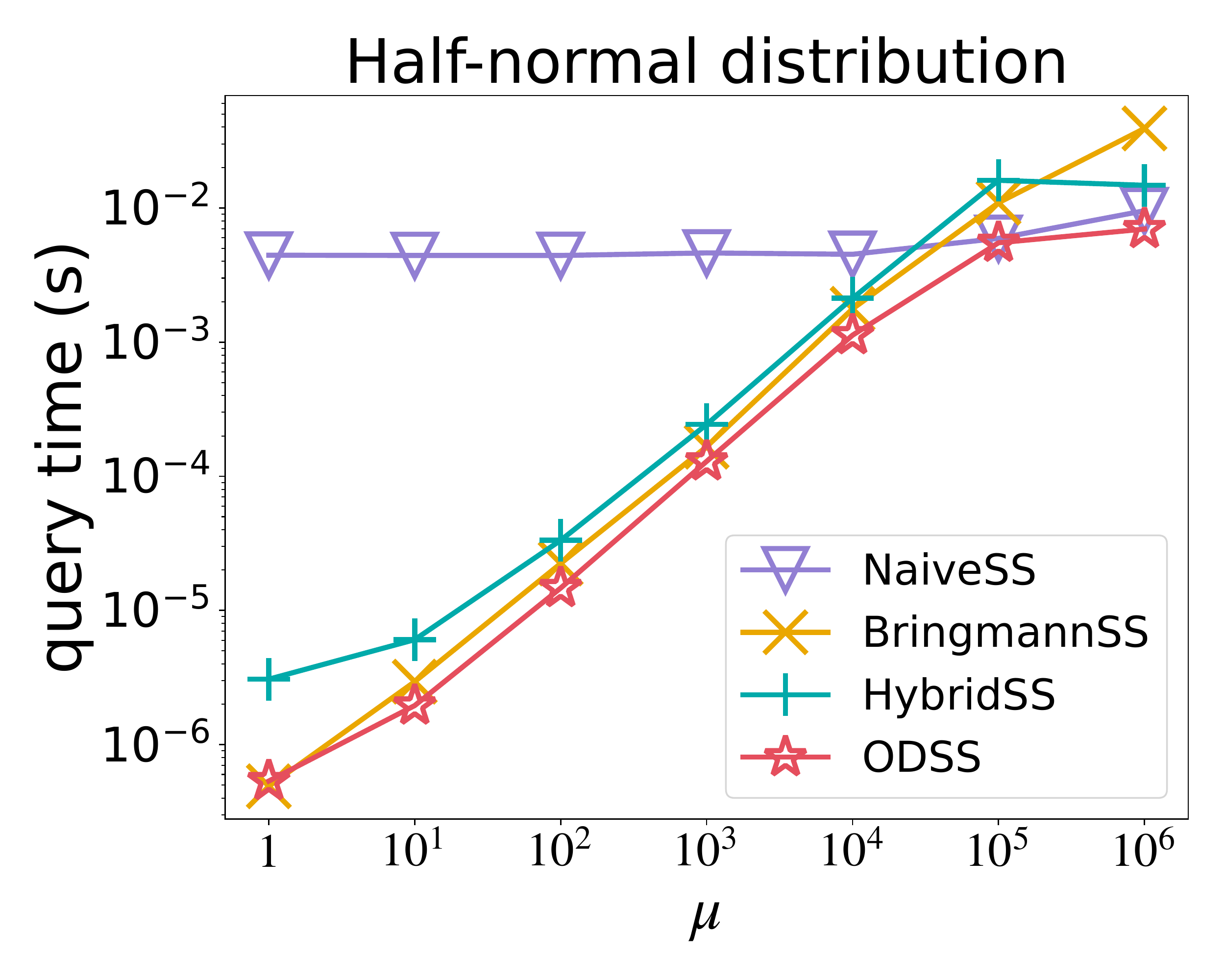}&
\hspace{-2mm} \includegraphics[width=44mm]{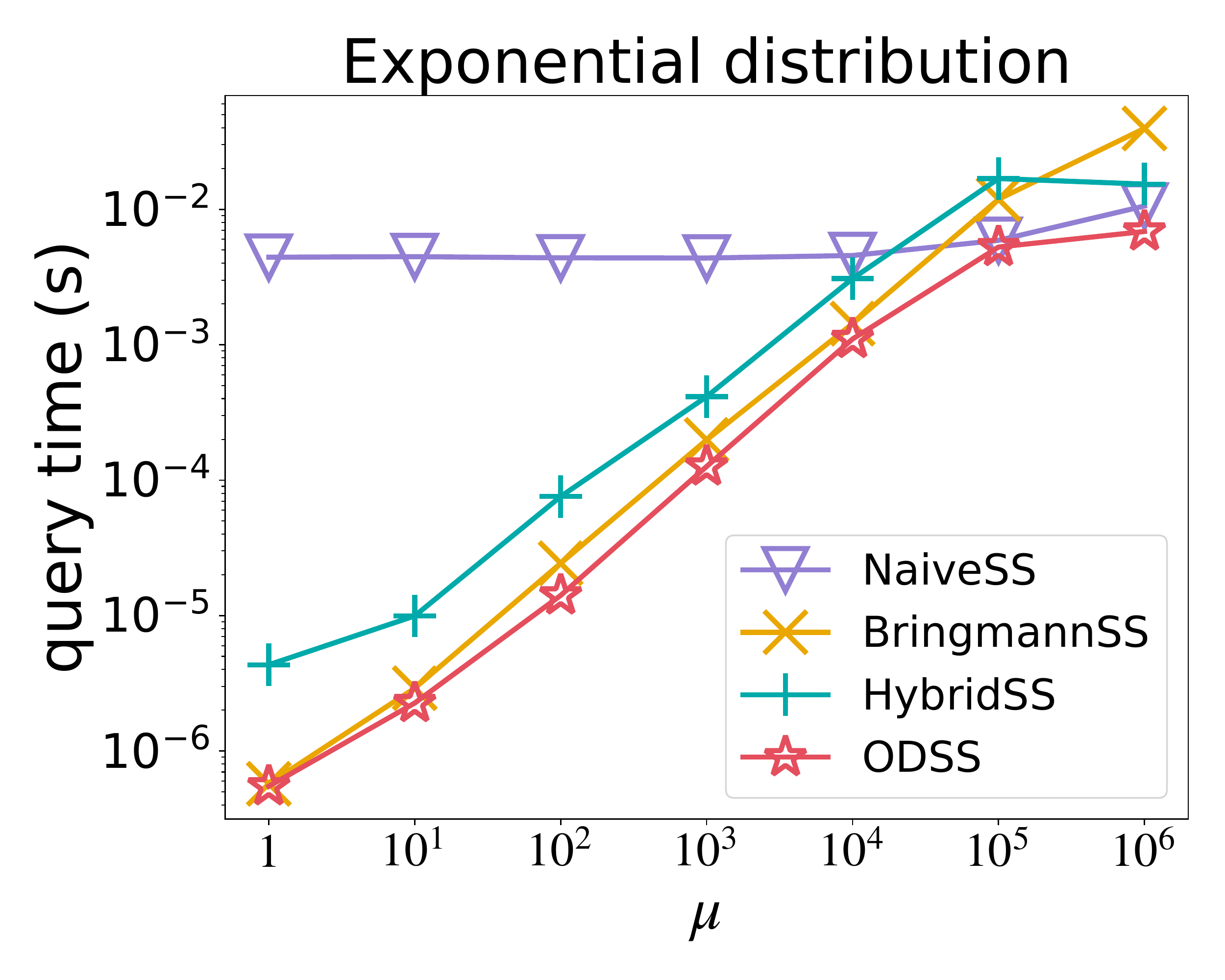} &
\hspace{-2mm} \includegraphics[width=44mm]{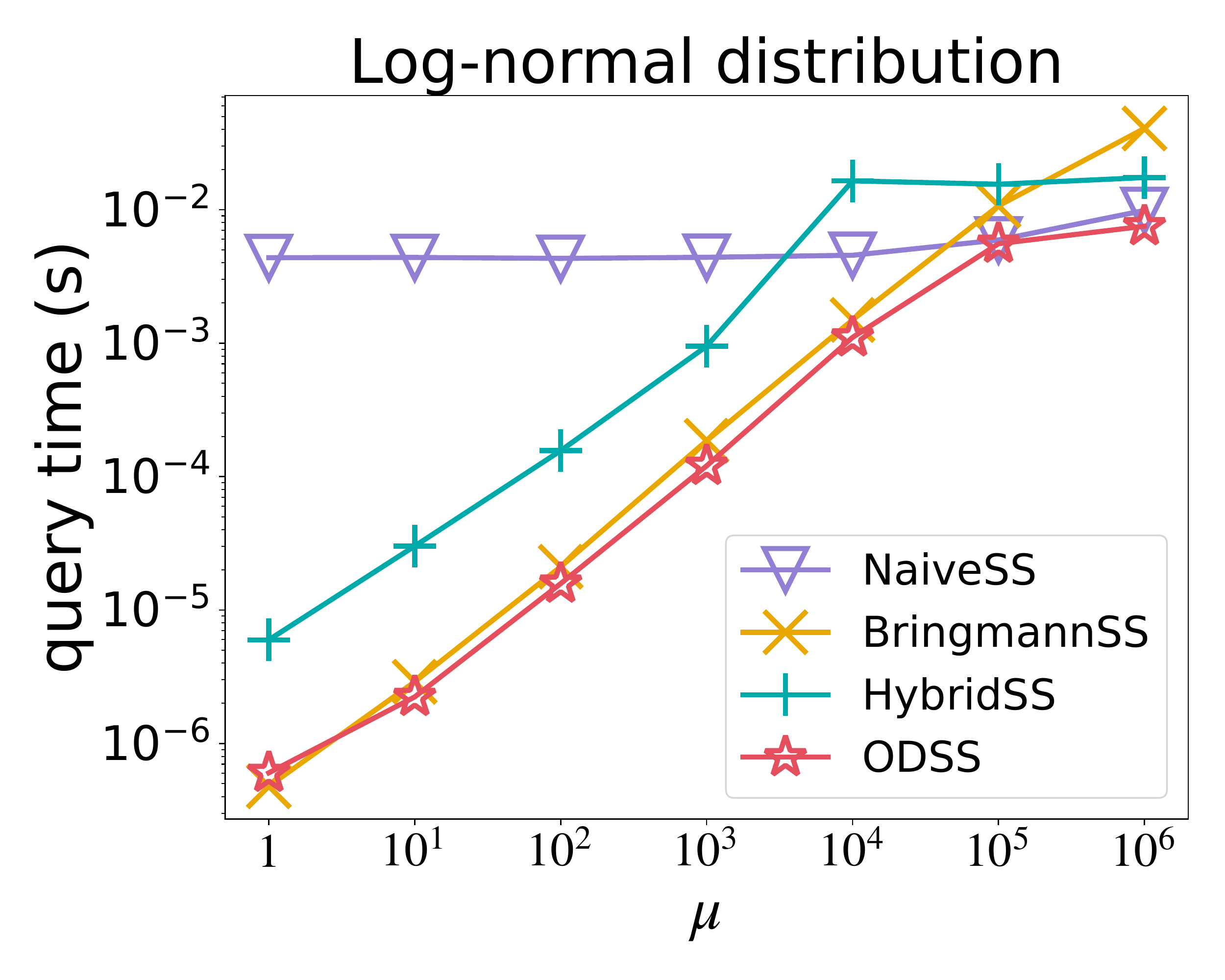} 
\hspace{-20mm}
\end{tabular}
\vspace{-6mm}
\caption{Varying $\mu$: query time (s) on distributions with different skewnesses. ($n=10^6$)}
\label{fig:query}
\end{minipage}

\begin{minipage}[t]{1\textwidth}
\centering
\vspace{1mm}
\begin{tabular}{cccc}
\hspace{-20mm} 
\includegraphics[width=44mm]{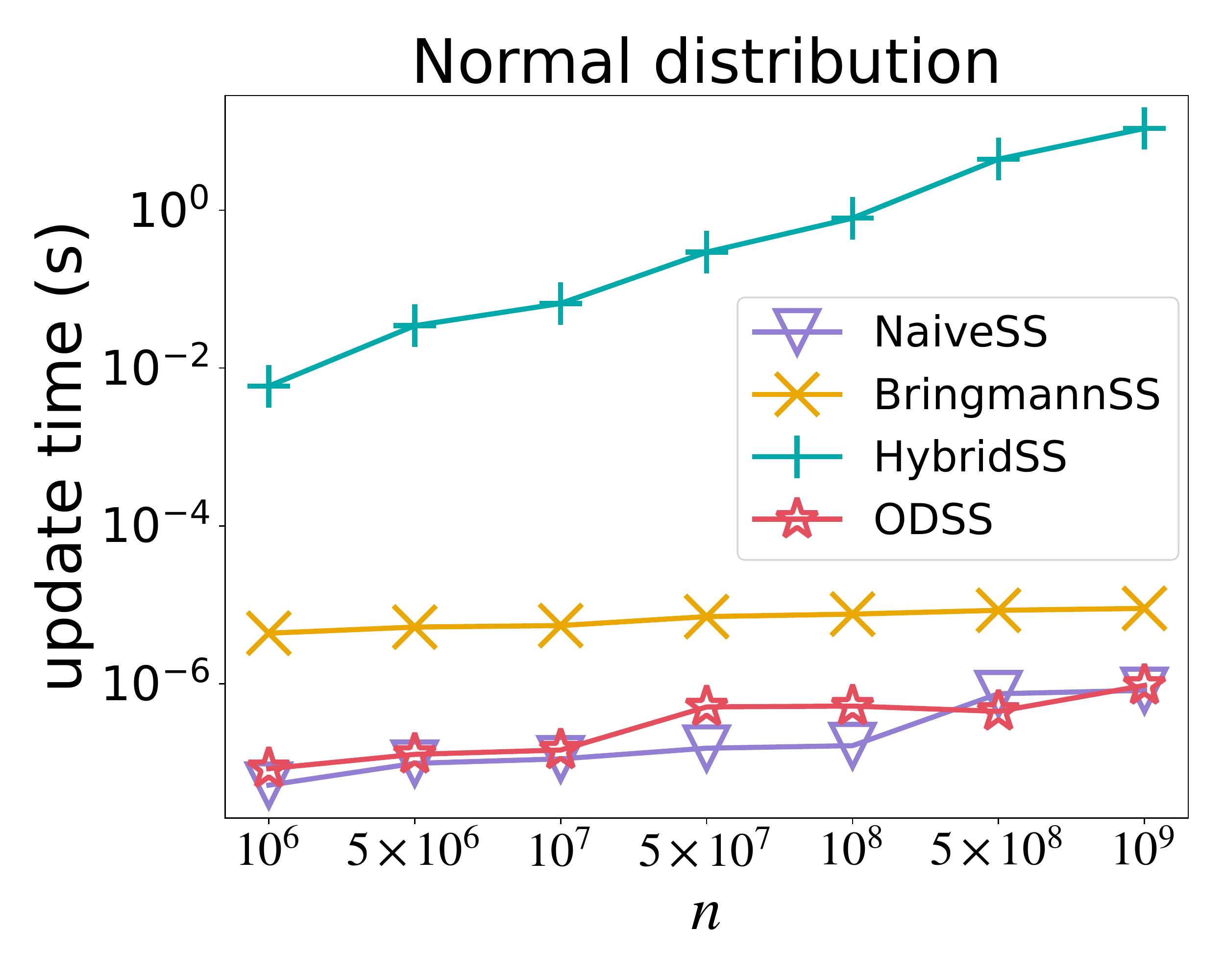}&
\hspace{-2mm} \includegraphics[width=44mm]{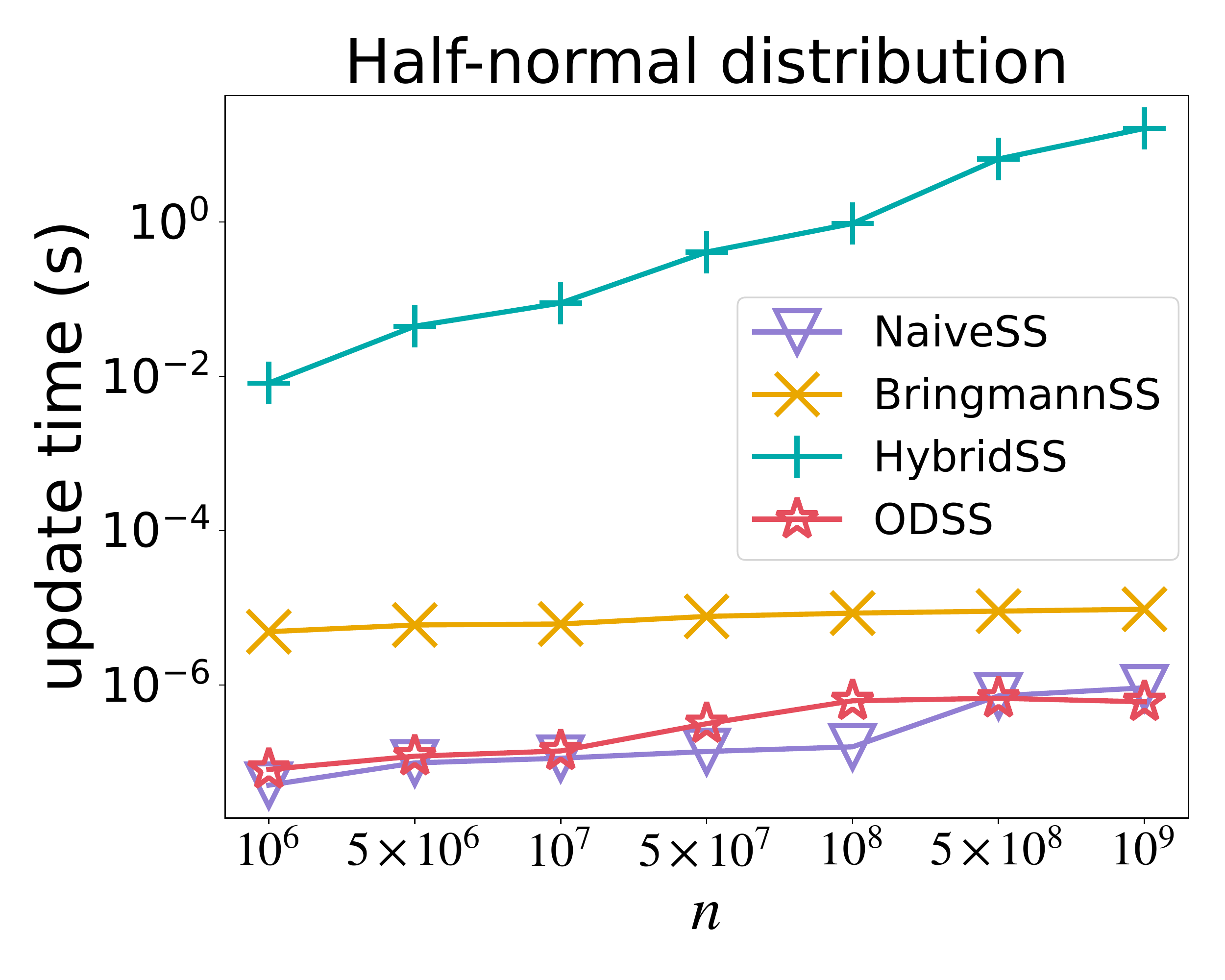}&
\hspace{-2mm} \includegraphics[width=44mm]{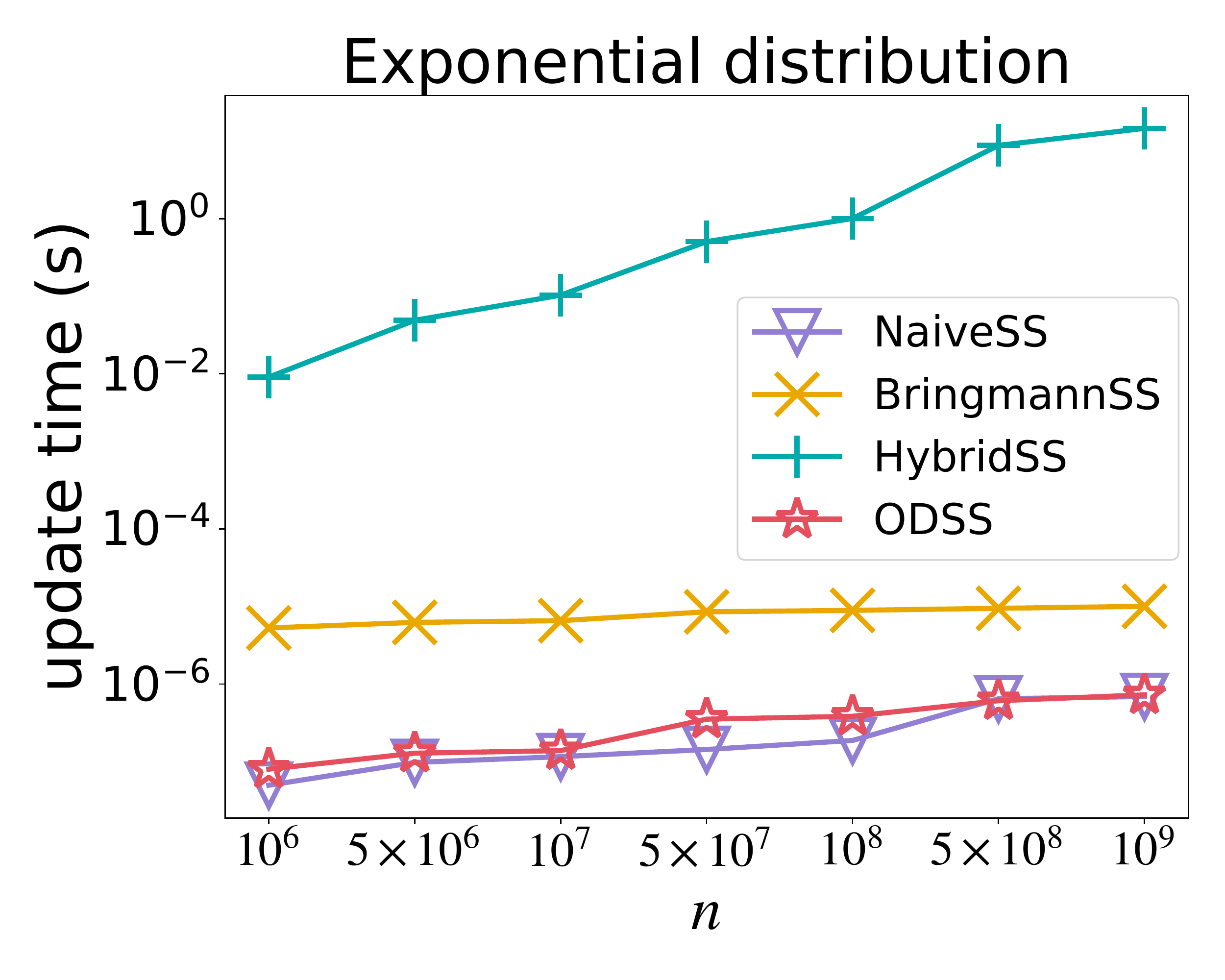} &
\hspace{-2mm} \includegraphics[width=44mm]{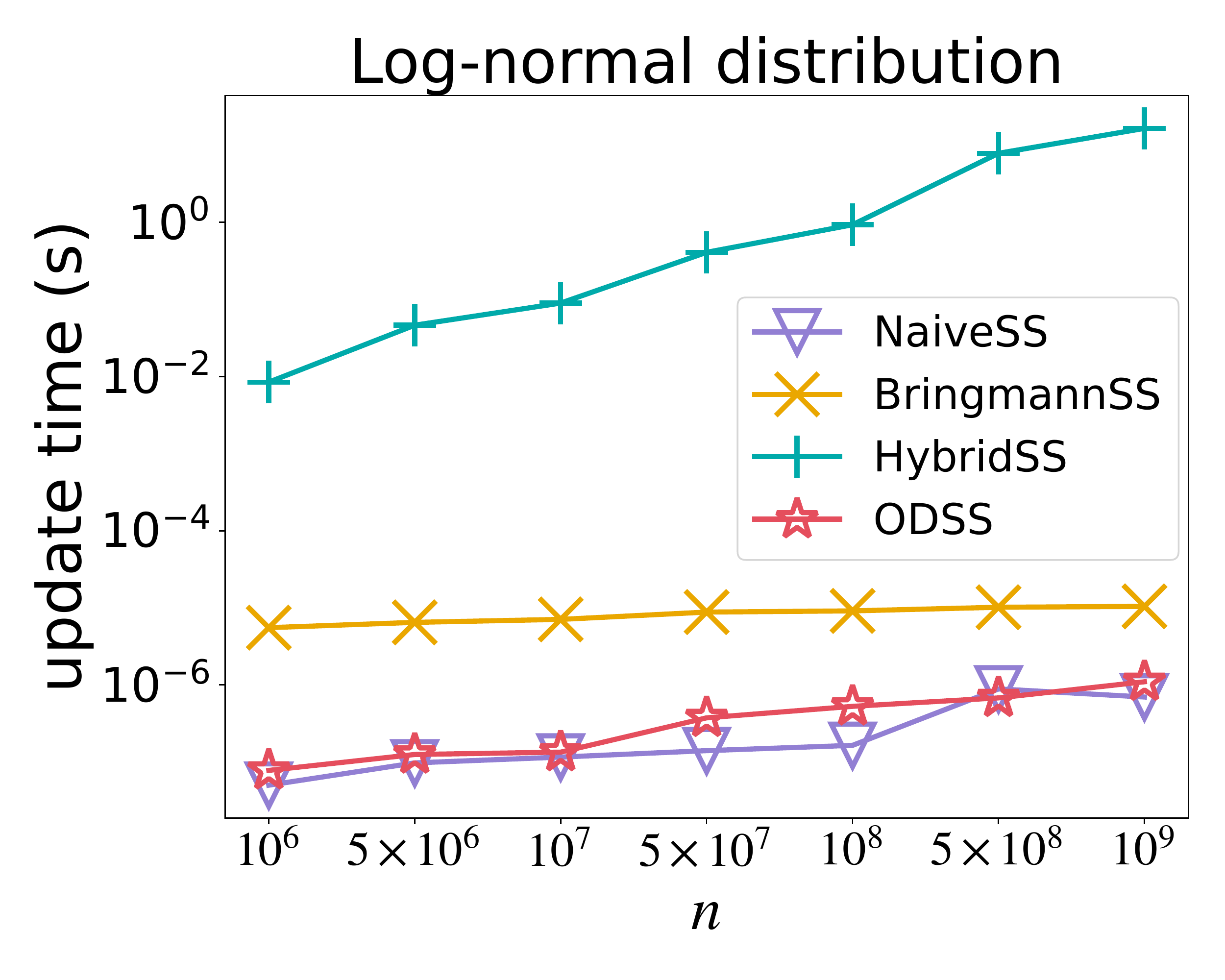} 
\hspace{-20mm}
\end{tabular}
\vspace{-6mm}
\caption{Varying $n$: update time (s) on distributions with different skewnesses.}
\label{fig:update}
\vspace{-3mm}
\end{minipage}
\end{figure*}

In this section, we experimentally evaluate the performance of our \odss against alternatives. Before the evaluation, we test the correctness of all algorithms through experiments, the results of which are deferred to Appendix~\ref{sec:app_exp}.

\header{\bf Environment.} We conduct all experiments on a machine with an Intel(R) Xeon(R) Silver 4114 CPU @ 2.20GHz CPU and 1007GB memory in Linux OS. All the methods are implemented in \verb|C++| and compiled by \verb|g++| with \verb|-O3| optimization.

\header{\bf Competitors.} We compared our \odss against the Naive method, the HybridSS method, and the BringmannSS method. For each method, we report the average time of $100$ queries as the query time, the average time of $10^5$ insertions and $10^5$ deletions as the update time.

\header{\bf The Distributions of Probabilities.} To examine the effectiveness of our \odss, we carefully set the probabilities of elements distributed with different \textit{skewnesses}. 
In particular, we choose the normal distribution, the half-normal distribution, the exponential distribution, and the log-normal distribution as the probability distribution of the elements. The normal distribution is a symmetric distribution with skewness as $0$. We set the mean of the normal distribution as $0$ and the variance as $10$. For the half-normal distribution, the \textit{probability density function (PDF)} is  $f(x)=\frac{\sqrt{2}}{\sigma \pi} \exp\big(-\frac{x^2}{2\sigma^2}\big)$ and the skewness of it is just below $1$. We set the $\sigma^2$ as $10$ for it. For the exponential distribution, the PDF is $f(x)=\lambda e^{-\lambda x}$ with the skewness as $2$, and we set $\lambda=1$. For the log-normal distribution, the PDF is $f(x)=\frac{1}{x\sigma \sqrt{2\pi}}\cdot \exp\big(-\frac{(\ln{x}-\mu)^2}{2\sigma^2}\big)$. By setting $\mu=0$ and $\sigma=\sqrt{\ln{2}}$, the skewness of the log-normal distribution is $4$. To generate the probabilities of elements, we first sample a value for each element with the distribution settings. To ensure that each value is greater than $0$, we subtract the minimum of all the generated values from each value. Then we re-scale the range of the values to $[0,1]$ and re-scale the sum of them to a specified $\mu$ according to the needs of various experiments. 

\header{\bf Query Time v.s. Update Time. } In light of the pressing needs for simultaneously achieving sample and update efficiency in the subset sampling problem, we draw the trade-off plots between \textit{query time} and \textit{update time} of each method in Figure~\ref{fig:tradeoff}. We set the number of elements $n=10^5$, and the sum of the probabilities is $\mu=1$. 
From Figure~\ref{fig:tradeoff}, we observe that \odss consistently achieves the best of both query time and update time on all datasets. The Naive method performs efficient updates, while it requires a large query time, which is $1000\times$ slower than \odss. 
The query time of the BringmannSS method and the HybridSS method is significantly smaller than the Naive method, while the update time is $100\times$, $10^4\times$ larger than our \odss, respectively. 
These results concur with our analysis for query time and update time. For update time, the Naive method and \odss are the best with only $O(1)$ time per element insertion/deletion, while the BringmannSS method needs $O(\log^2 n)$ time and the HybridSS methods need $O(n)$ time. For query time, the BringmannSS method, and our \odss are the best with $O(1+\mu)$ time. The HybridSS method needs $O\left(1+n\sqrt{\min\left\{\bar{p},1-\bar{p}\right\}}\right)$ query time, and the mean of the probabilities $\Bar{p}$ is so small here, and thus the HybridSS performs better than the Naive method, which needs $O(n)$ time in any case.
 
\header{\bf Effectiveness of Queries. } To further examine the query time of these algorithms, we vary $\mu$, the sum of the probabilities of elements, and show the query time in Figure~\ref{fig:query}. We set $n=10^6$, and vary $\mu$ in $\{1,10,10^2,10^3,10^4,10^5,10^6\}$. From Figure~\ref{fig:query}, we observe that with the increase of $\mu$, the query time of all algorithms except the Naive method becomes larger. This concurs with the theoretical analysis that the query time of the BringmannSS method and \odss is $O(1+\mu)$, and the increase of the mean $\Bar{p}$ of the probabilities incurs the increase of the query time of the HybridSS method. Additionally, we note that when $\mu=n=10^6$, the query time of the BringmannSS method is larger than that of the Naive method due to the complicated index structure. \odss costs the least query time with any $\mu$ in this experiment.

\header{\bf Effectiveness of Updates. } To further examine the update time of these algorithms, we set $\mu=1$ and vary $n$ in $\{10^6,5\times 10^6,10^7, 5\times 10^7,10^8,5\times 10^8,10^9\}$. Figure~\ref{fig:update} shows the update time of the algorithms on the four distributions. We observe that the Naive method and \odss consistently outperform other algorithms on all the distributions. The update time of the BringmannSS method is $10\times \sim 50\times$ larger than that of the Naive method and our \odss. The update time has little change with the increase of $n$ since the $O(\log^2 n)$ term increases mildly when $n$ varies from $10^6$ to $10^9$. The update of the HybridSS method becomes slower with the increase of $n$, which concurs with its update time complexity $O(n)$.

\section{Empirical Study}\label{sec:application}

In this section, we apply our \odss to a concrete application: Influence Maximization (IM) of evolving graphs. We will first briefly introduce the IM problem and the dynamic IM problem as follows.

With the rapid growth of online social networks, the study of information diffusion in networks has garnered significant attention. Among the various research areas in information diffusion, Influence Maximization (IM) has emerged as a critical problem with practical applications in viral marketing~\cite{domingos2001mining}, network monitoring~\cite{leskovec2007cost}, and social recommendation~\cite{ye2012exploring}. The objective of IM is to identify a set of seed users in a social network that can maximize the spread of influence.

In this context, we consider a social graph $\G=(\V,\E)$, where $\V$ represents the set of nodes and $\E$ represents the set of edges. Each edge $e=(u,v)\in \E$ is associated with a propagation probability $ p(u,v) \in [0,1]$. To capture the stochastic process of influence diffusion on $\G$, we define a specific \textit{diffusion model} $\M$. The \textit{influence spread} of a set of nodes $S$, denoted as $\sigma_{\G,\M}(S)$, is the expected number of users influenced by $S$ under the diffusion model $\M$.

\begin{definition}[Influence Maximization~\cite{kempe2003maximizing}]
Given a graph $\G$, a diffusion model $\M$, and a positive integer $k$, the Influence Maximization problem aims to select a set $S_k$ of $k$ nodes from $\G$ as the seed set to maximize the influence spread $\sigma_{\G,M}(S_k)$, i.e., $S_k=\arg \max_{S:|S|\le k} \sigma_{\G,\M}(S)$.
\end{definition}

\header{\bf The IC Model. }In this work, we focus on a widely adopted model, the \textit{Independent Cascade (IC)} model. Under the IC model, a seed set $S\in \V$ spreads its influence as follows. At timestep $0$, all nodes in $S$ are activated. Each node $u$ that is activated at timestep $t-1$ has only ONE chance to activate each of its inactive outgoing neighbors $v$ with probability $p(u,v)$. After the time step $t$, $u$ stays active and do not activate any nodes again. The spreading process terminates when no more active nodes can activate other nodes. We observe that the process by which an active node $u$ activates its outgoing neighbors is exactly a subset sampling problem. Each outgoing neighbor $v$ of $u$ is independently sampled with probability $p(u,v)$. 

\header{\bf Static IM. } It has been proved that IM is NP-hard under the Independent Cascade (IC) model~\cite{kempe2003maximizing}. Due to the theoretical hardness of IM, extensive research has tried to design efficient IM algorithms for a few decades. According to ~\cite{li2018influence}, the existing algorithms can be classified into three categories: the sketch-based~\cite {borgs2014maximizing,tang2014influence,tang2018online} solutions, the simulation-based solutions~\cite{kempe2003maximizing,goyal2011celf++,leskovec2007cost,zhou2015upper,wang2010community}, and the proxy-based solutions~\cite{chen2009efficient,jung2012irie,cheng2014imrank,liu2014influence}. The sketch-based solution and the simulation-based solution both solve a huge number of subset sampling problems.

\begin{figure*}[t]
\begin{minipage}[t]{1\textwidth}
\centering
\begin{tabular}{cccc}
\hspace{-20mm} 
\includegraphics[width=43mm]{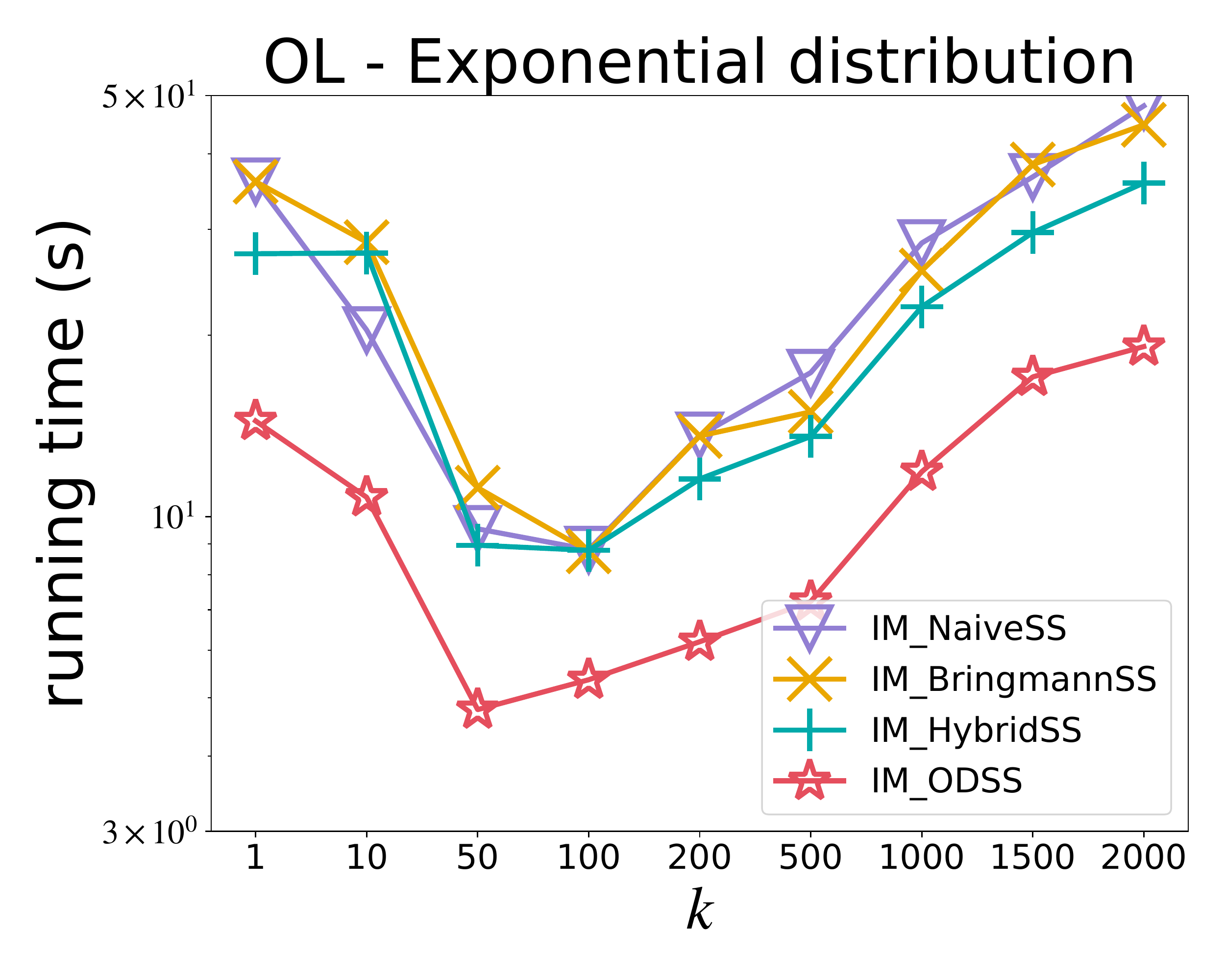}&
\hspace{-2mm} \includegraphics[width=43mm]{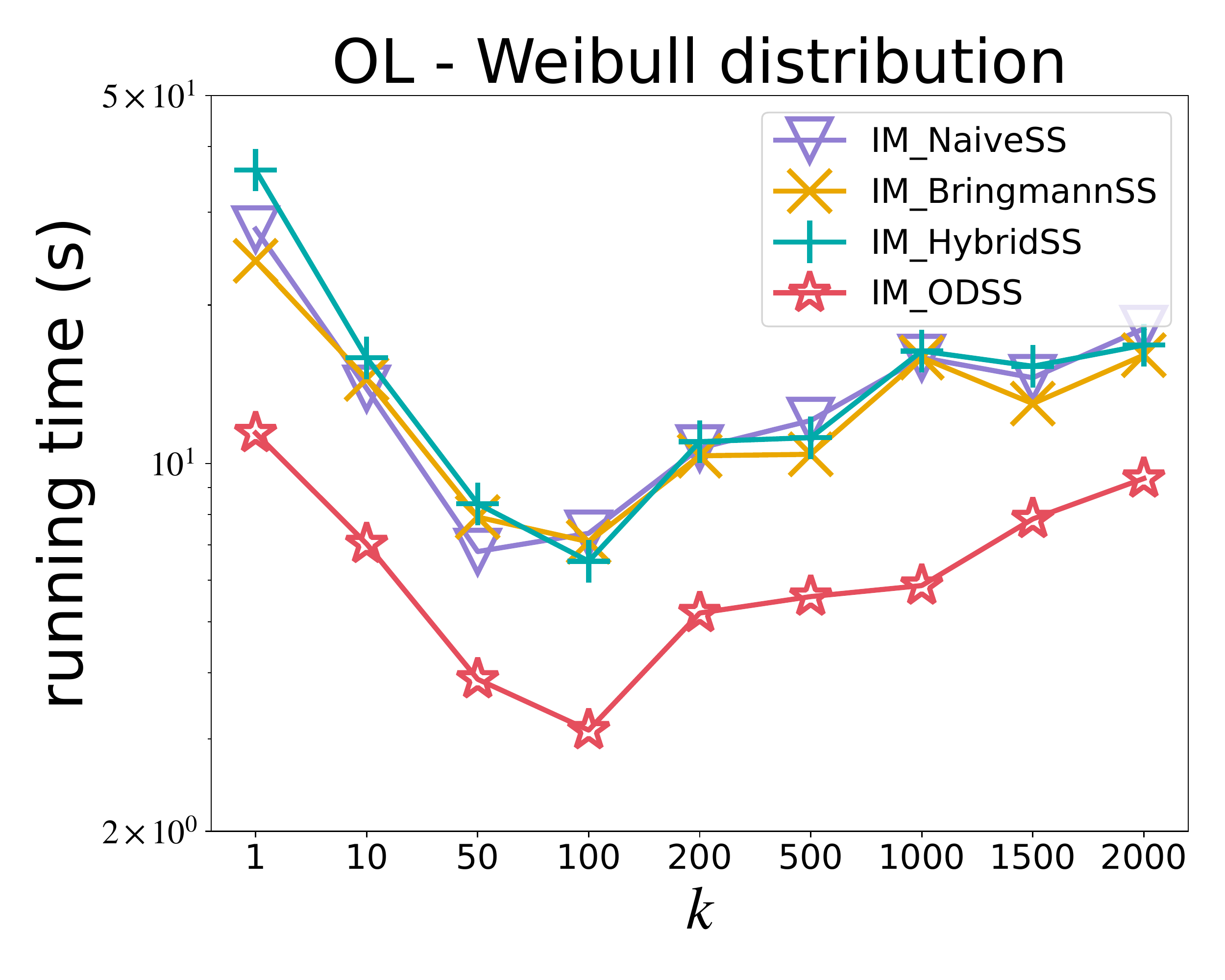} &
\hspace{-2mm} \includegraphics[width=43mm]{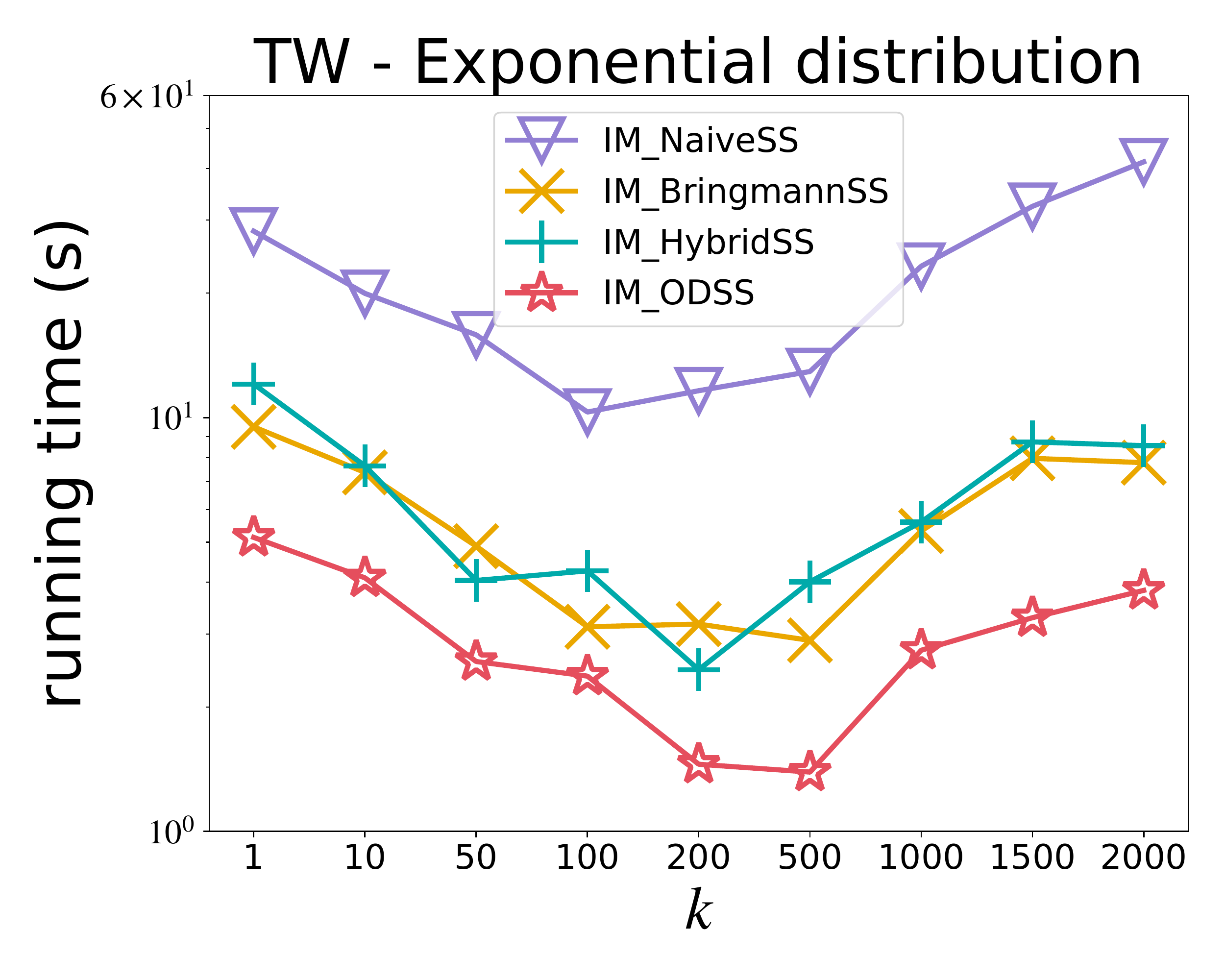} &
\hspace{-2mm} \includegraphics[width=43mm]{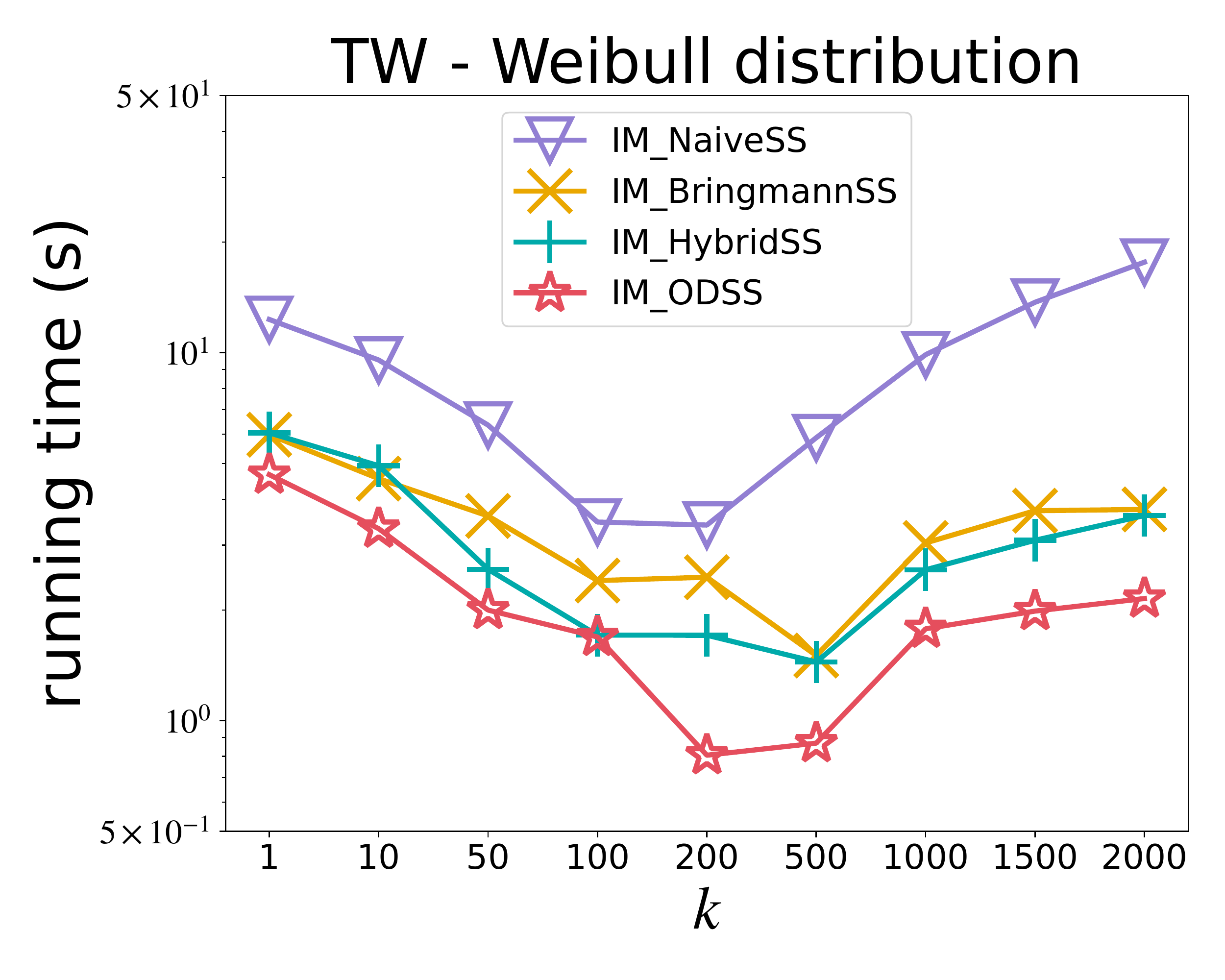}
\hspace{-20mm}
\end{tabular}
\vspace{-5mm}
\caption{Running time of dynamic IM algorithms based on various subset sampling structures.}
\label{fig:rrset}
\vspace{2mm}
\end{minipage}

\begin{minipage}[t]{1\textwidth}
\centering
\begin{tabular}{cccc}
\hspace{-20mm} 
\includegraphics[width=43mm]{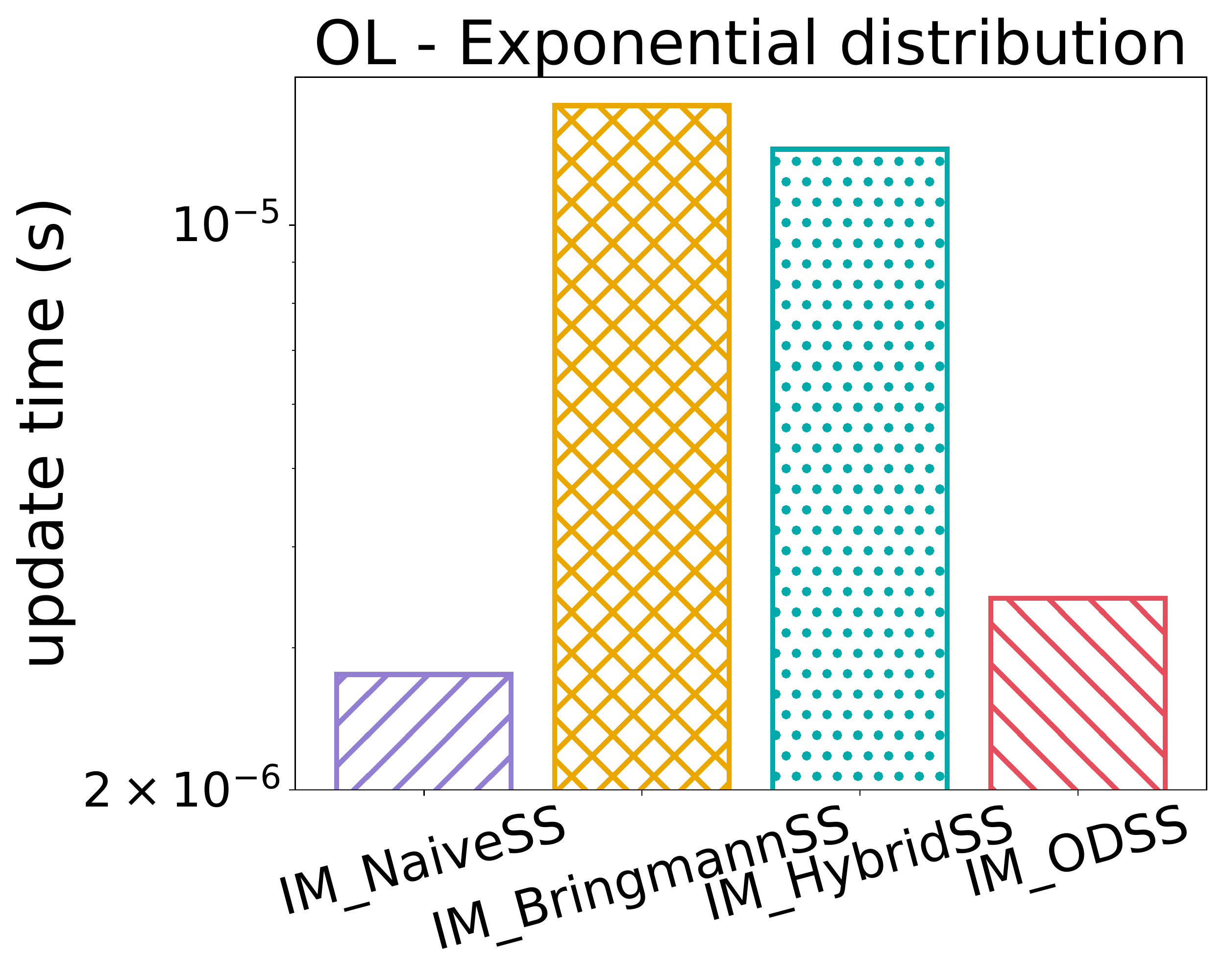}&
\hspace{-2mm} \includegraphics[width=43mm]{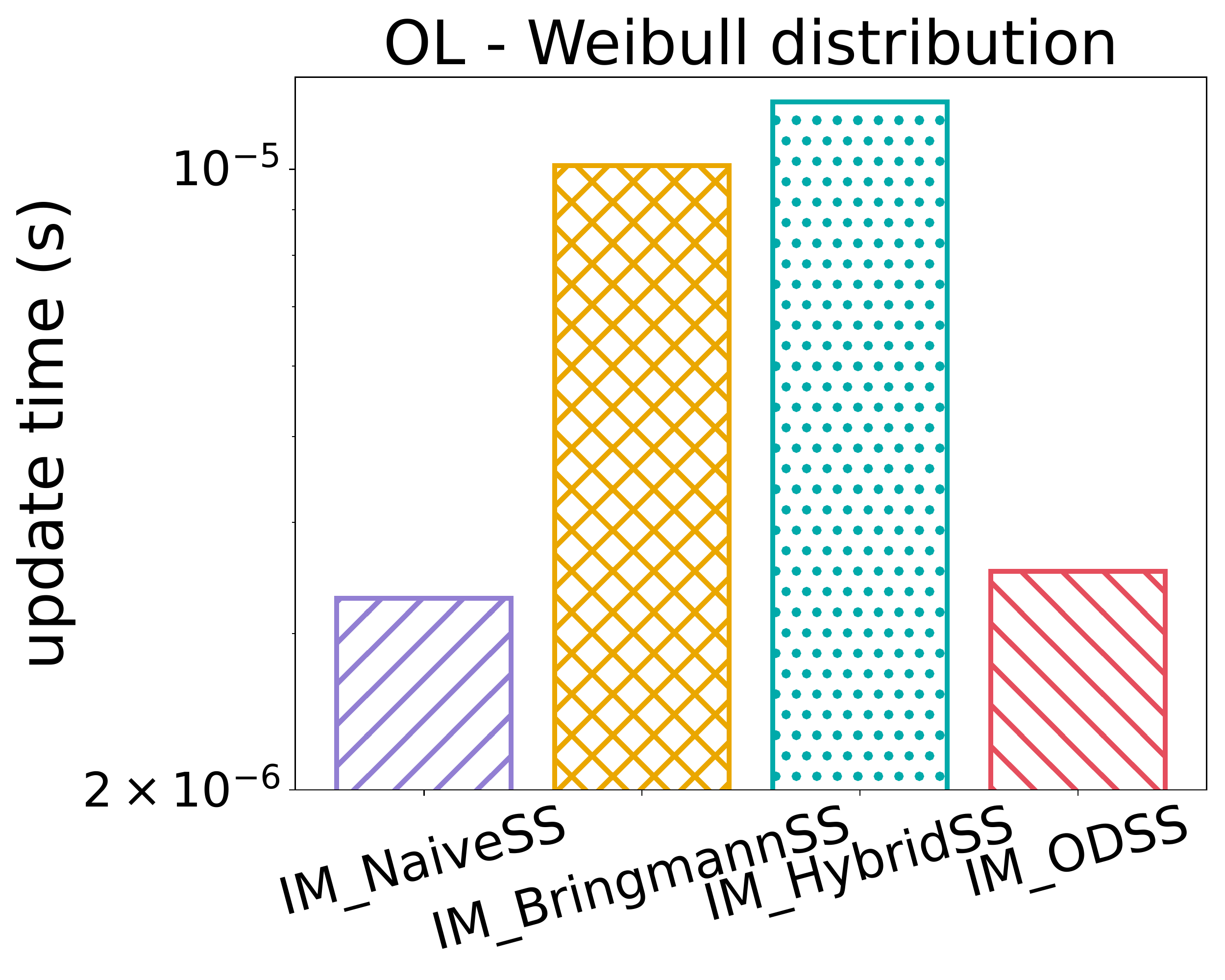} &
\hspace{-2mm} \includegraphics[width=43mm]{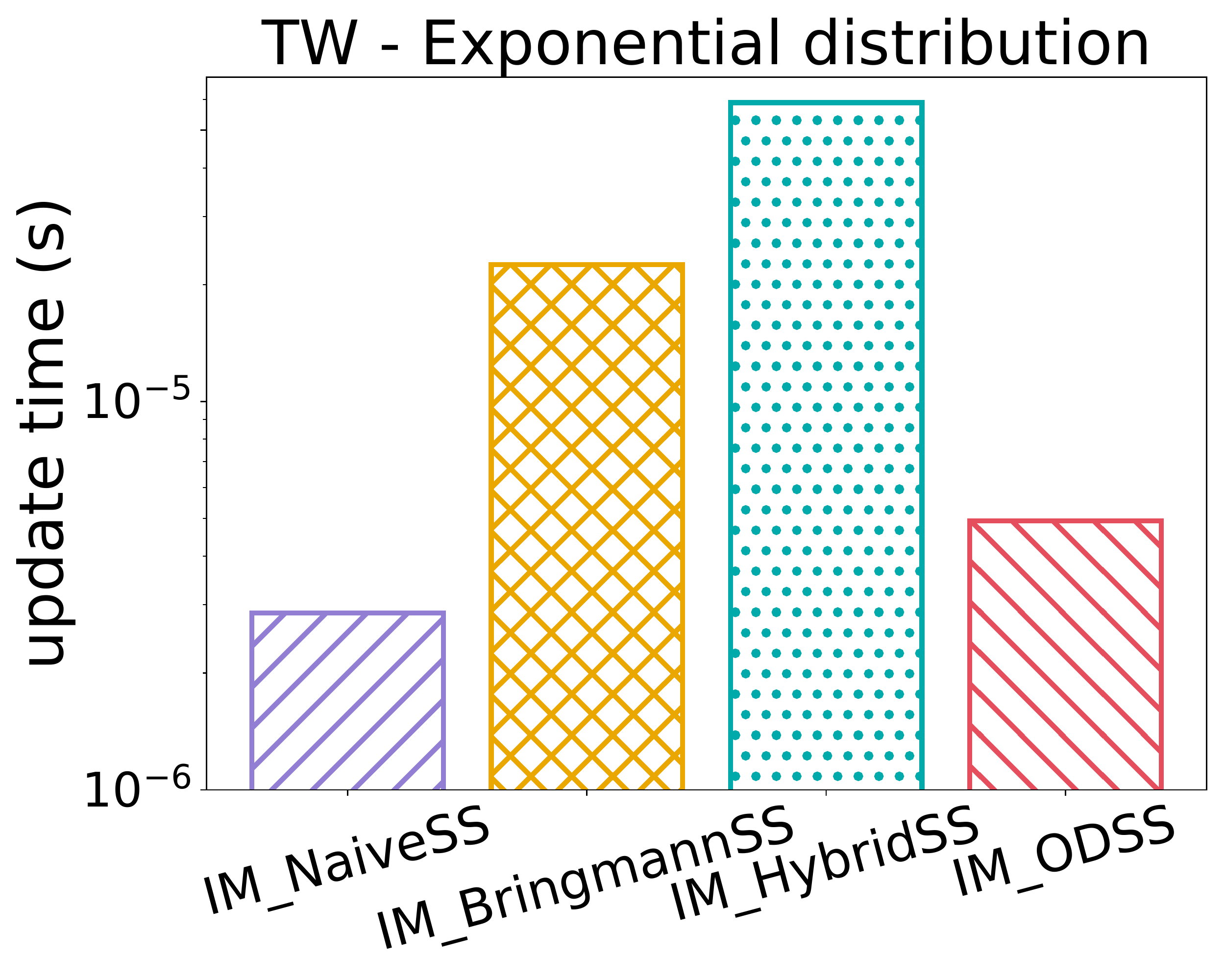} &
\hspace{-2mm} \includegraphics[width=43mm]{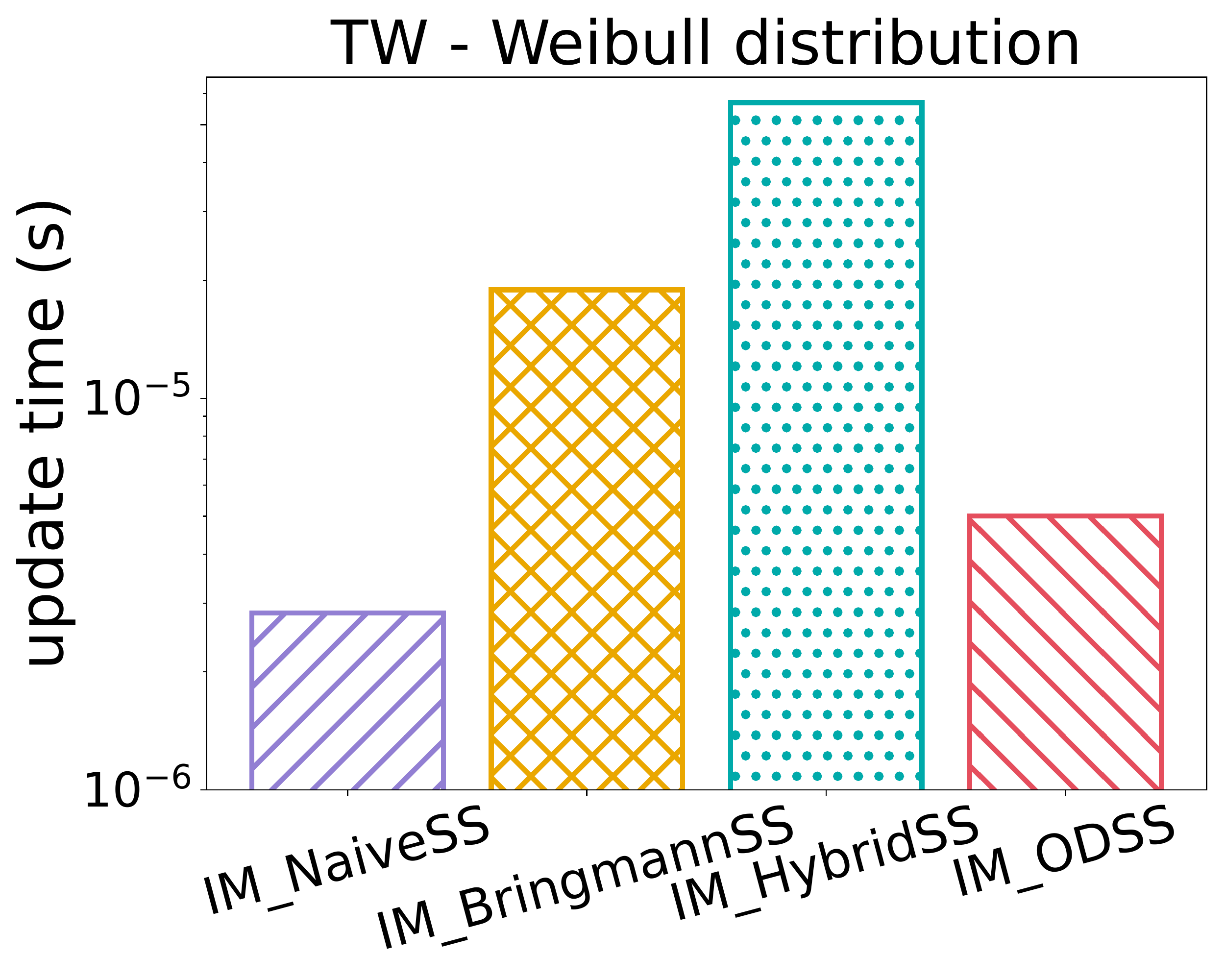}
\hspace{-20mm}
\end{tabular}
\vspace{-5mm}
\caption{Update time of dynamic IM algorithms based on various subset sampling structures.}
\label{fig:rrupd}
\vspace{-2mm}
\end{minipage}
\end{figure*}

\header{\bf Dynamic IM. }Considering the highly dynamic nature of social influence, the \textit{dynamic IM} problem has received much attention in the past decade. Dynamic IM hopes to support real-time influential users tracking on evolving social networks. 
According to Peng~\cite{peng2021dynamic}, in the fully dynamic network model where users can join in or leave the networks, no algorithms can achieve any meaningful approximation guarantee. That is, re-running an IM algorithm upon each update can achieve the lower bound of the running time. However, the existing static IM algorithms focus on static social networks and do not support evolving networks. 
This motivates us that by replacing the subset sampling module in the static IM algorithms with the structure of \odss, we can obtain a new dynamic IM algorithm for the fully dynamic model. The query time of each subset sampling problem in the new IM algorithm is reduced to $O(1+\mu)$. Meanwhile, since our \odss costs $O(1)$ time for each update operation, the new IM algorithms can support insertions/deletions of user nodes or modifications of propagation probabilities in constant time. 

In the following, we will apply our \odss and alternatives to static RR-sketch solutions to obtain new dynamic IM algorithms for the fully dynamic model in Section~\ref{sec:sketch}. We also examine how our \odss improves the complexities of simulation-based solutions. Due to the simplicity of the simulation-based solutions, we defer the details in Appendix~\ref{sec:simu}. 


\begin{table}[t]
\centering
\caption{Datasets.}
\vspace{-4mm}
\begin{small}
\begin{tabular}{lrrr}
\toprule
{{\bf Dataset}}& {{\bf $\boldsymbol{n}$}} & {{\bf $\boldsymbol{m}$}} & {{\bf  $\boldsymbol{m/n}$}}\\ \midrule
{Orkut-Links} &{$3,072,441$}&{$117,185,083$}&{$38$}\\
{Twitter} & {$41,652,230$} & { $1,468,365,182$} & {$35$}\\
\bottomrule
\end{tabular}
\end{small}
\label{tbl:datasets}
\vspace{-4mm}
\end{table}


\subsection{Applying \odss to RR-Sketch Solutions}\label{sec:sketch}

The Reverse Reachable Sketch (RR-Sketch) solutions are the current mainstream of the sketch-based solutions for IM~\cite{li2018influence,tang2014influence,borgs2014maximizing}. The solutions are based on the concept of \textit{random Reverse Reachable (RR) set}. 
Generating a random RR set starts at a uniformly selected node $s$ from $\V$. 
At timestep $0$, $s$ is activated. Each node $v$ that is activated at timestep $t-1$ tries to reversely activate each ingoing neighbor $u$ with probability $p(u,v)$ at timestep $t$. After timestep $t$, $v$ stays active and no longer activates any nodes. The process ends when no more nodes can be reversely activated. The set of the active nodes is a random RR set, denoted as $R$. 
The activation is a subset sampling problem with the ingoing neighbors of a node as elements. After generating a sufficient number of random RR sets, a greedy algorithm is applied to select a seed set~\cite{borgs2014maximizing}.

We notice that Guo et al.~\cite{guo2020influence} propose a framework called SUBSIM (Subset Sampling with Influence Maximization). SUBSIM modifies the subset sampling module of the algorithm OPIM-C~\cite{tang2018online} and thus accelerates the generation of random RR sets. 
However, SUBSIM focuses on static IM problems and can not support the evolving graphs. 
Motivated by SUBSIM, we can obtain various dynamic IM algorithms for the fully dynamic model by replacing the subset sampling module with various dynamic subset sampling structures (our \odss and alternatives). We conduct experiments for these dynamic IM algorithms to evaluate the running time for IM and the update time for edge insertions/deletions.

The experiments are conducted on two real-world graphs, Orkut-Links (OL) and Twitter (TW). The two graphs are publicly available at ~\cite{snapnets}. We give the summary of the two graphs in Table~\ref{tbl:datasets}. Following previous studies~\cite{guo2020influence,tang2018online}, we test the case when the probabilities of edges follow two skewed distributions: exponential distribution and Weibull distribution. For exponential distribution, We set $\lambda=1$. For Weibull distribution, the PDF is $f(x)=\frac{a}{b} \cdot (\frac{x}{b})^{(a-1)}\cdot e^{-(x/b)^a}$. The parameters $a$ and $b$ are drawn uniformly from $[0,10]$ for each edge. For each node, the sum of the probabilities of its outgoing neighbors is scaled to $1$. 

To fully explore the efficiency of the algorithms, we vary the size $k$ of the seed set in $\{1,10,50,100,200,500,1000,2000\}$ in the experiments. We repeat each IM algorithm to generate the seed set $5$ times and report the average running time. To simulate the evolution of social networks, we uniformly choose $10^6$ edges from each graph and report the average time for the insertions and deletions of these $10^6$ edges. 
Figure~\ref{fig:rrset} presents the running time of the dynamic IM algorithms based on various subset sampling structures. IM\_ODSS, the IM algorithm based on our \odss, outperforms alternatives in all tested graphs and all seed sizes. In particular, the running time of IM\_ODSS is $10\times$ smaller than the IM\_NaiveSS on the TW graph when $k\ge 500$. Figure~\ref{fig:rrupd} presents the update time of the algorithms. IM\_ODSS performs per update in less than $4\times 10^{-6}$ s. IM\_BringmanSS and IM\_HybridSS both suffer the long update time. In particular, IM\_HybridSS is $10\times$ slower than IM\_ODSS on the TW graph.

\section{Conclusion}

Subset sampling is a fundamental problem in both data mining and theoretical computer science. Its dynamic version finds various applications in Influence Maximization, Graph Neural Networks, and Computational Epidemiology. This paper proposes \odss, the first optimal dynamic subset sampling algorithm. We present a theoretical analysis to demonstrate the optimal complexities of \odss. We also conduct extensive experiments to evaluate the performance of \odss and give an empirical study on Influence Maximization.
\bibliographystyle{plain}
\balance
\bibliography{paper}
\appendix
\clearpage
\section{Supplementary Experiments} \label{sec:app}

\subsection{Implementation Details}\label{sec:app_structure}
\header{\bf Maintaining Arrays for Groups. } To achieve the update operations, we maintain an array for each group so that accessing the element at a specified location can be done in constant time. 
We also maintain an array for storing the group index $k$ and the position $j$ for each element $x_i$, indicating that $x_i$ is the $j$-th element in $G_k$. 
Let ${\sf Gidx}(x_i)$ be the group index of $x_i$ and ${\sf Gpos}(x_i)$ be the position of $x_i$ in the group. When inserting an element into a group, we append it to the end of the array. 
When it comes to deletion, we remove $x_i$ from the ${\sf Gidx}(x_i)$-th group by replacing the value at position ${\sf Gpos}(x_i)$ with the value at the last position in the group and then delete the last cell. 
That is, we move the last element to the position of $x_i$, and then the array contracts by one cell. 
Consequently, the insertion or deletion of an element within a group can be accomplished in constant time. 
Note that each update operation can be resolved into a constant number of element insertions or deletions within a group. 
Consequently, every update operation can be performed in constant time. 
The memory cost of a group is directly proportional to the number of elements it contains. Consequently, the memory cost of level $0$ is $O(n)$ and the memory cost of level $1$ is $O(\log n)$.

\header{\bf Maintaining Dynamic Arrays. }Note that the size of each group will grow and shrink with insertions and deletions of elements within the group. To support the changes of the groups and bound the total memory in the meantime, we can implement the arrays using the \textit{doubling technique}. 
Specifically, for each array $A$, we maintain an additional array $A+$. Denote the size of $A$ as $c$, initially $c=2$. Let $n'$ be the number of elements in $A$. The size of $A+$ is initialized as $2c$. 
In the beginning, $A+$ is empty. When $n'$ starts to exceed $c/2$, each time we insert a new element into $A$, the new element is inserted into $A+$ as well and one element contained in $A$ is copied to $A+$. Thus, when $A$ is full, $A+$ contains all elements in $A$. Then, we release the memory space of $A$, make $A+$ as the new $A$, and create a new $A+$ of size $4c$. When $n'$ becomes smaller than $\frac{c}{4}$, we delete $A+$ and contract $A$ by $\frac{c}{2}$. Then we create a new $A+$ of size $c$. 
Note that the time cost for each insertion/deletion is still bounded by $O(1)$, and the memory space of $A+$ is asymptotically the same as that of $A$. Thus, the total memory space of level $0$ and level $1$ only depends on the number of elements at the level, which is $O(n)$ and $O(\log n)$, respectively.

\header{\bf Managing the Dynamic Number of Groups. } 
The number of groups at both level $0$ and level $1$ may dynamically change over time as the number of elements $n$ increases or decreases. 
We consider how to maintain $O(1)$ update time in this scenario. 
To achieve this, we make the assumption that the upper bound of $n$ is known in advance. 
With the assumption, we maintain the lookup table with $m=\lceil \log (\lceil \log n \rceil +1) \rceil+1$ where $n$ is the upper bound of the number of elements over time. 
Next, we address the challenge of maintaining the correct group partitions when the number of groups, denoted as $K$, varies. 
When $K$ is reduced to $K-1$, we conceptually combine the $K$-th and $(K-1)$-th groups into a new group, $G_{K-1}$. 
This involves recalculating $p(G_{K-1})$ based on the new group size and setting a flag to indicate the combination. 
For empirical implementations, we can just move two elements from the old group $G_K$ to the old group $G_{K-1}$ in each sample or update operation. 
Before the number of groups is further reduced by one, all elements have been moved into $G_{K-1}$. 
To handle an increase in $K$, we implicitly maintain the partition within $G_K$. The implicit group ${G'}_{K}$ only consists of elements with probabilities no greater than $2^{-K+1}$ and the implicit group ${G'}_{K+1}$ contains the other elements. When $K$ is increased to $K+1$, the implicit ${G'}_{K+1}$ departs from $G_K$ and becomes the new last group $G_{K+1}$. During subsequent query or update operations, we move an element from the new group $G_{K+1}$ into either the new implicit group ${G'}_{K+1}$ or the new implicit group ${G'}_{K+2}$ based on its probability. This process ensures that the partition is completed before the number of groups reaches $K+2$. Consequently, the correct group maintenance is achieved in constant time.

\begin{figure*}[t]
\begin{minipage}[t]{1\textwidth}
\centering
\vspace{-2mm}
\begin{tabular}{cccc}
\hspace{-20mm} 
\includegraphics[width=40mm]{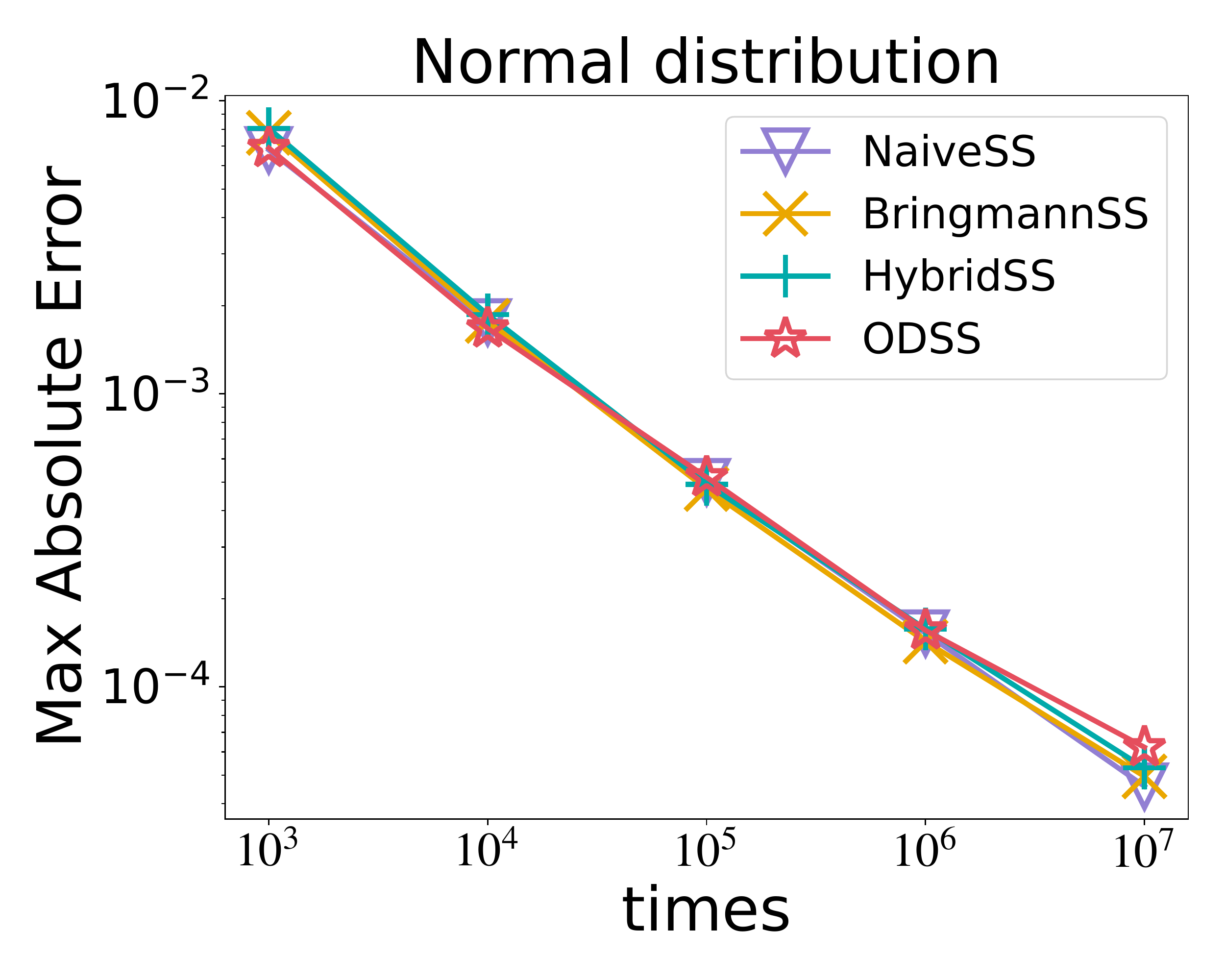} &
\hspace{-2mm} \includegraphics[width=40mm]{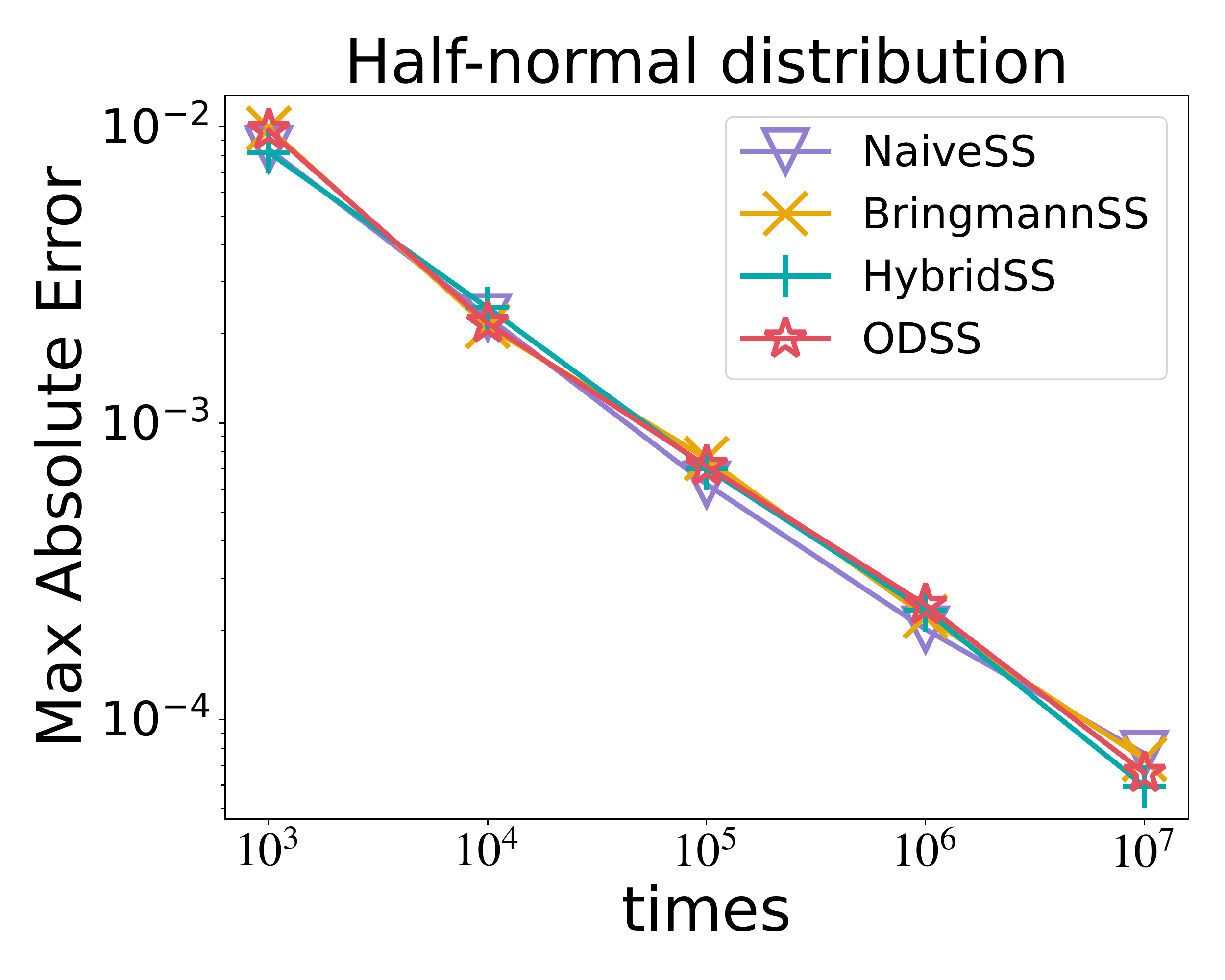} &
\hspace{-2mm} \includegraphics[width=40mm]{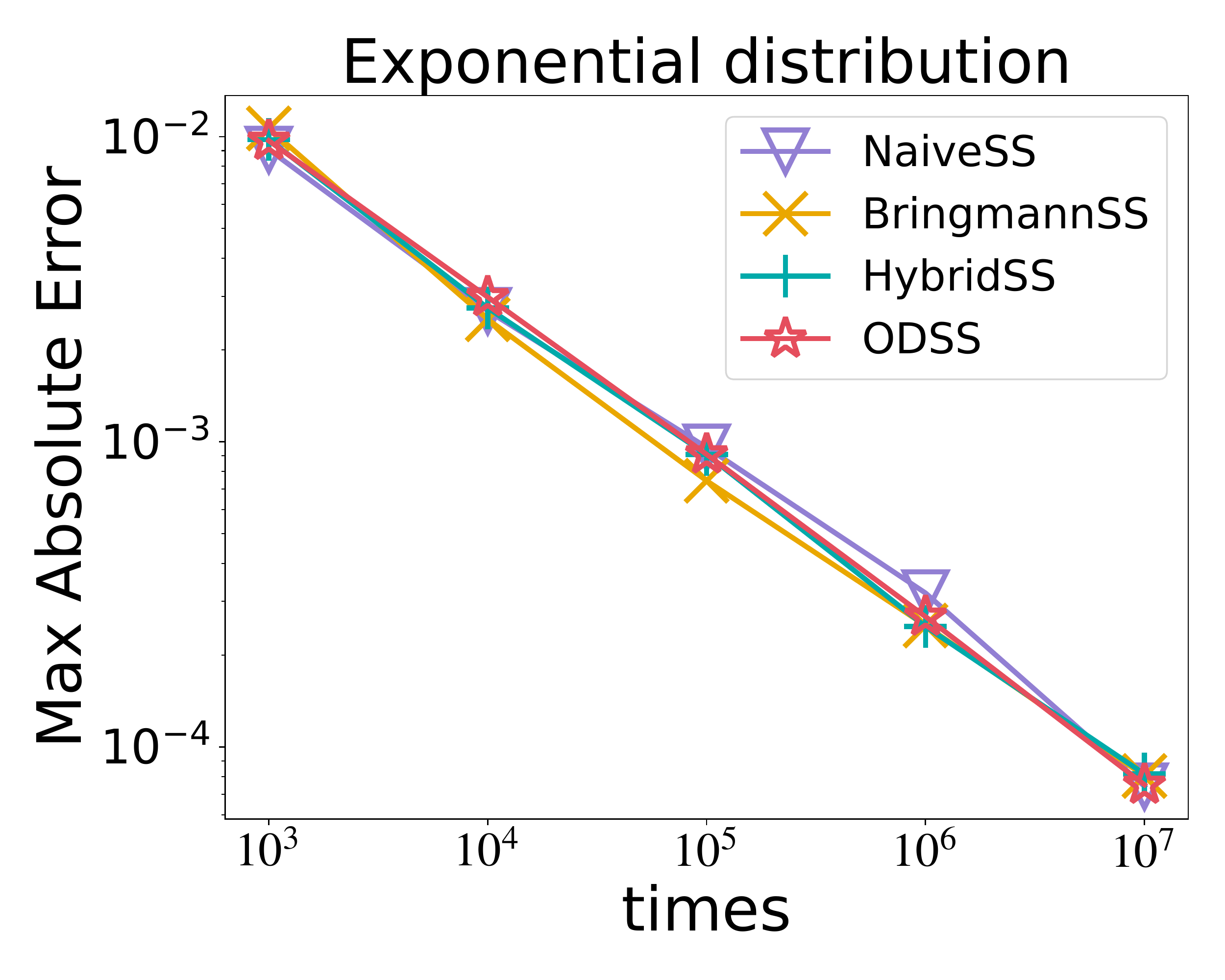} &
\hspace{-2mm} \includegraphics[width=40mm]{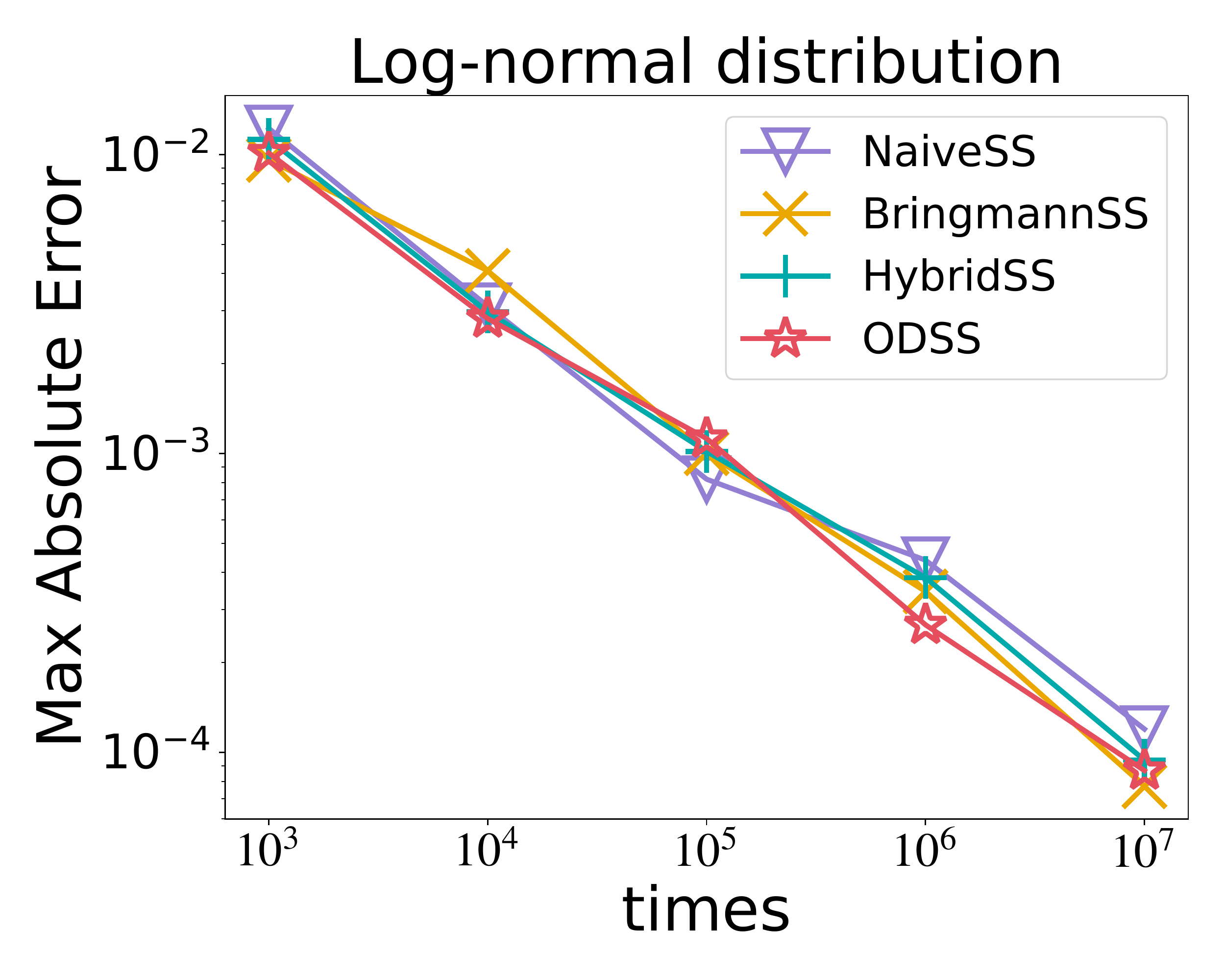}
\hspace{-20mm}
\end{tabular}
\vspace{-5mm}
\caption{Max absolute error v.s. repeat times on distributions with different skewnesses. }
\label{fig:err}
\vspace{-1mm}
\end{minipage}
\end{figure*}

\begin{figure*}[t]
\begin{minipage}[t]{1\textwidth}
\centering
\vspace{-2mm}
\begin{tabular}{cccc}
\hspace{-20mm} 
\includegraphics[width=40mm]{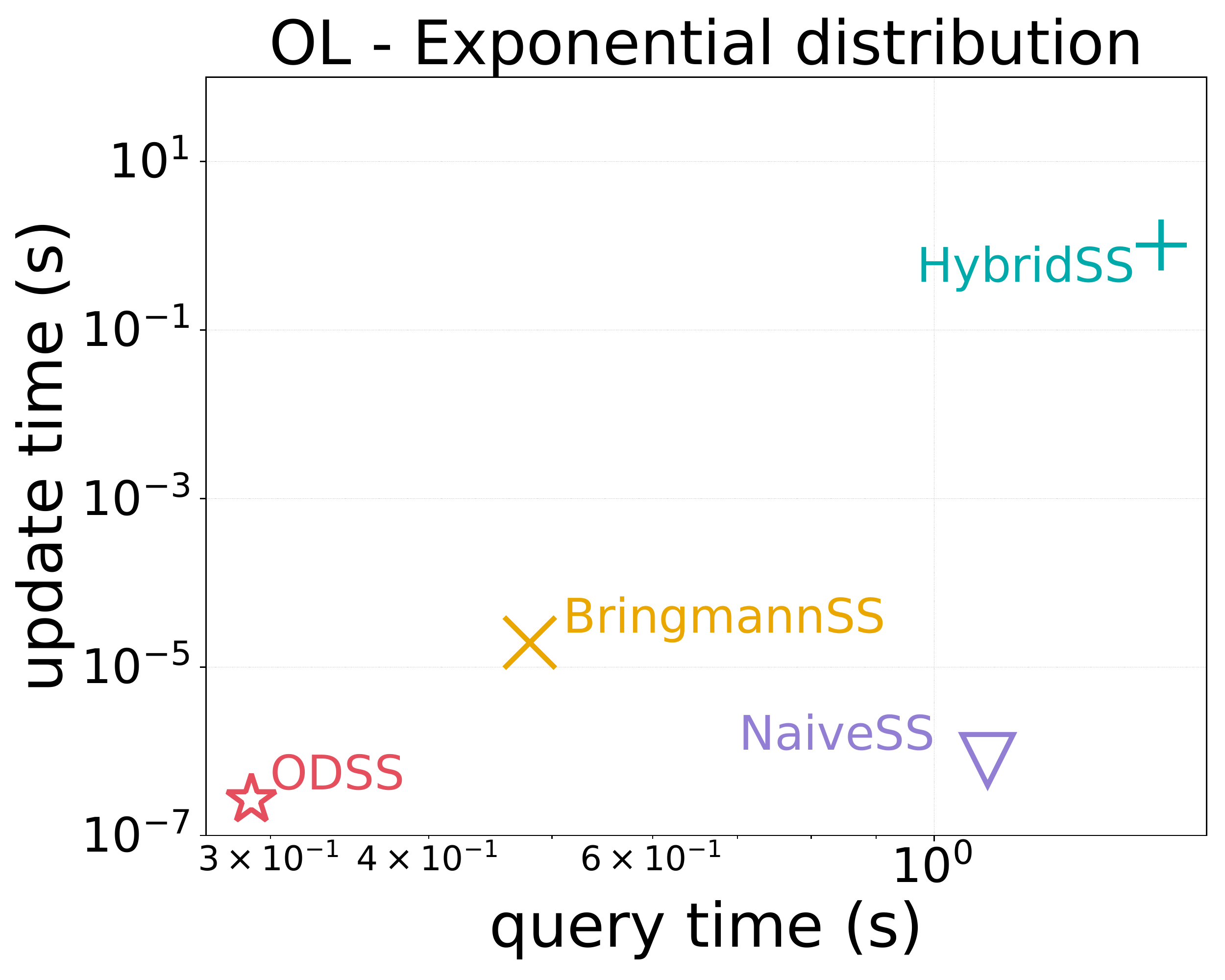}&
\hspace{-2mm} \includegraphics[width=40mm]{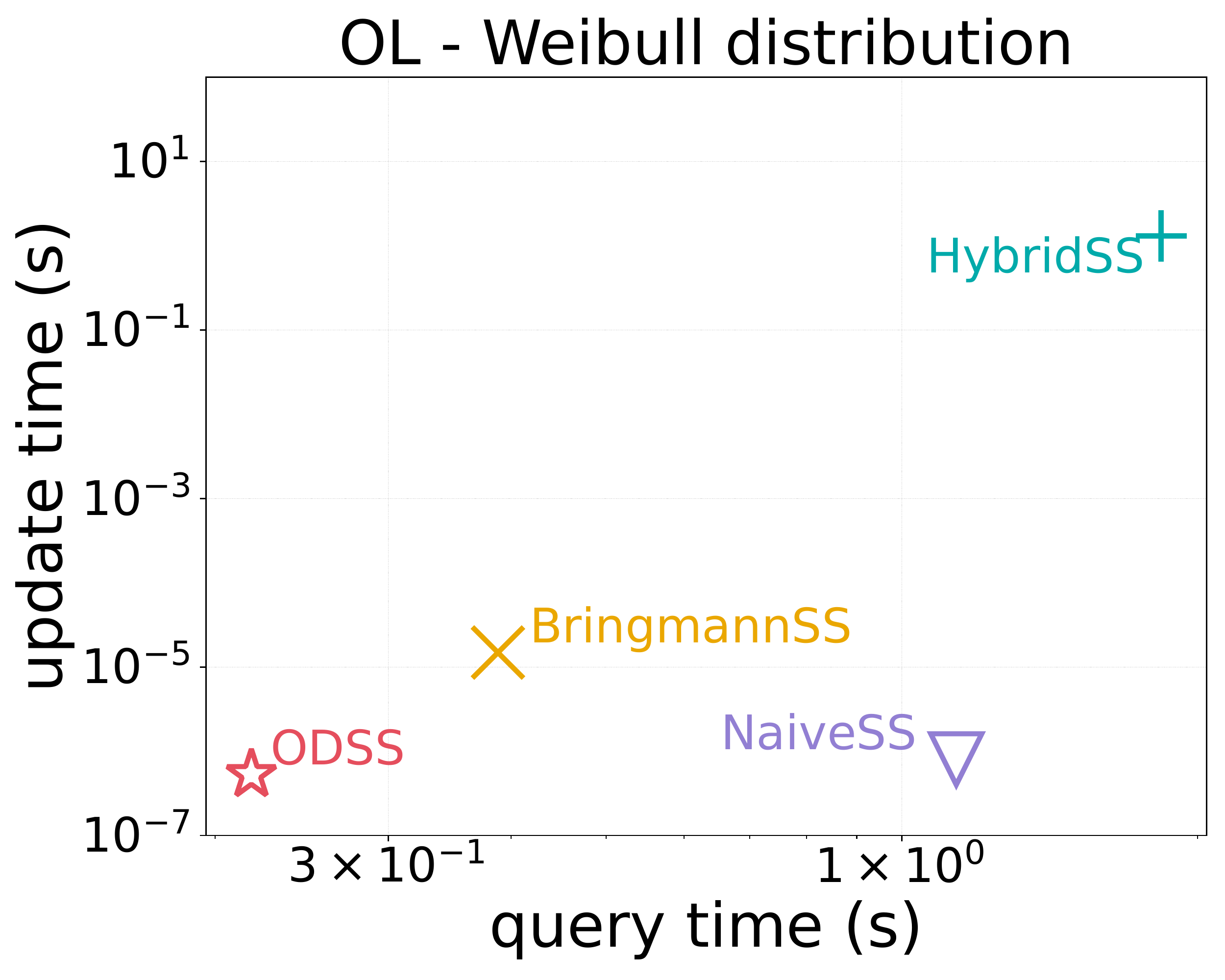} &
\hspace{-2mm} \includegraphics[width=40mm]{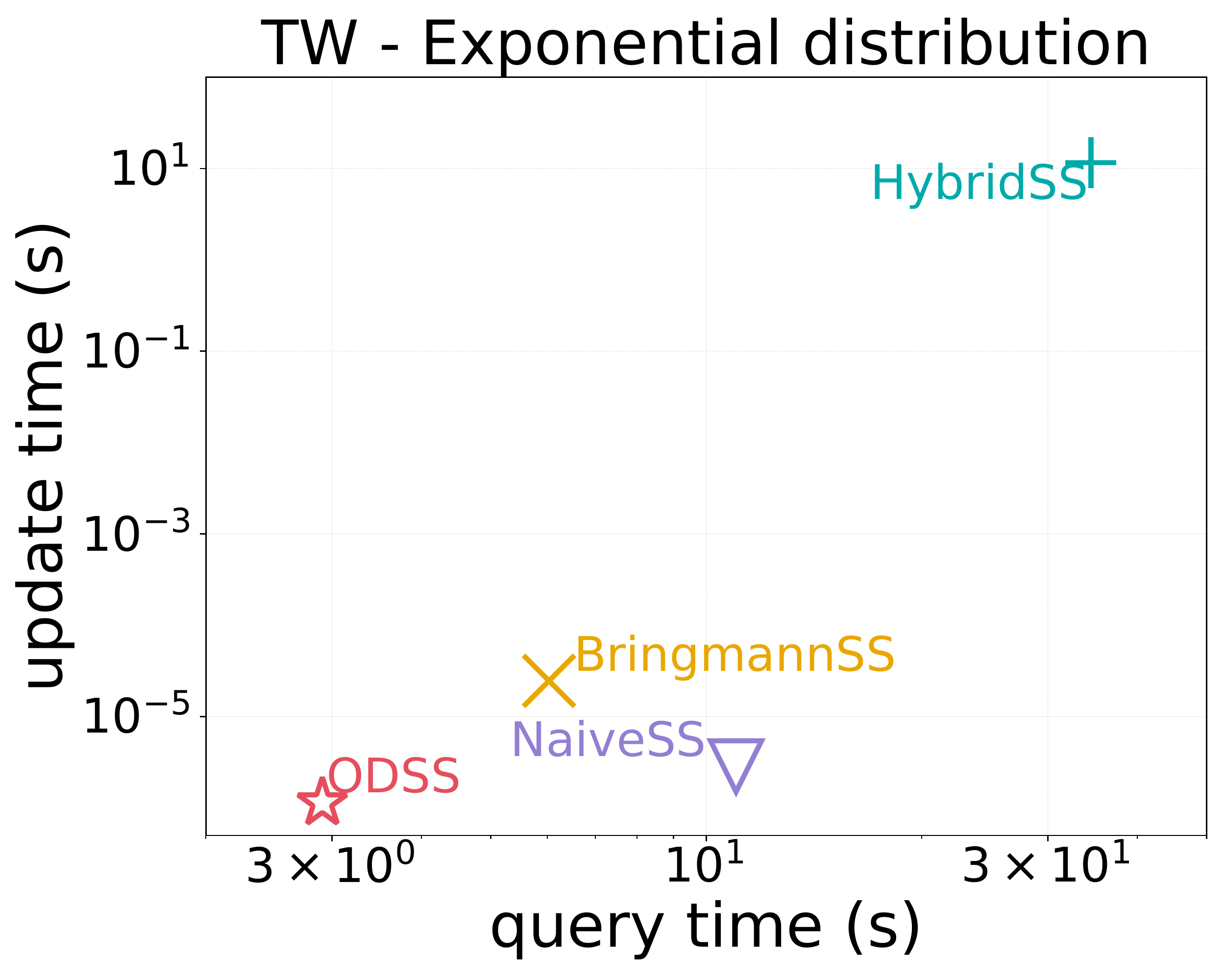} &
\hspace{-2mm} \includegraphics[width=40mm]{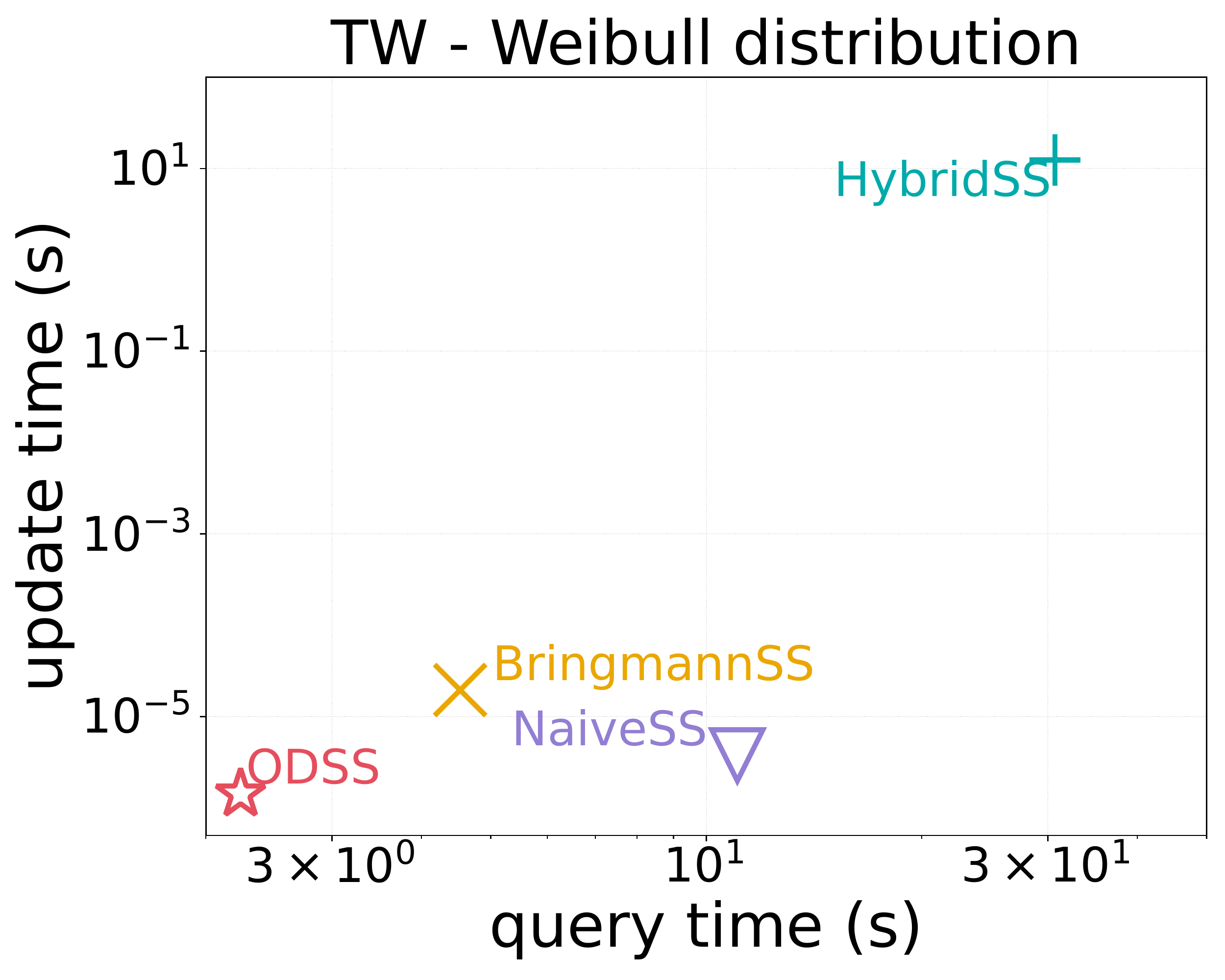}
\hspace{-20mm}
\end{tabular}
\vspace{-5mm}
\caption{Query time v.s. update time for generating possible worlds. }
\label{fig:world}
\vspace{-1mm}
\end{minipage}
\end{figure*}
\subsection{Experiments on Empirical Error} \label{sec:app_exp}
We conduct experiments to empirically check the correctness of our \odss and alternatives. According to Law of large numbers, the empirical probability of each element obtained from a large number of queries will be close to the true probability and tends to become closer to the true probability as more queries are performed. Thus, we repeat queries, calculate the empirical probability $\hat{p}(x)$ for each element $x$, and report the maximum absolute error, $\max_{x\in S}|\hat{p}(x)-p(x)|$. Specifically, we set the number of elements as $n=10^5$, the sum of the probabilities of elements $\mu=100$, and the number of update operations as $1000$. We first perform $1000$ element insertions and then perform $1000$ element deletions and repeat the subset sampling queries for $\{10^3, 10^4, 10^5, 10^6, 10^7\}$ times. Thus, the update algorithm and the query algorithm are both tested. Figure~\ref{fig:err} shows the maximum absolute error of all algorithms with different repeat times. We observe that the maximum absolute error decreases with increasing repeat times on all algorithms, which implies the correctness of these algorithms.

\subsection{ODSS for Dynamic Influence Maximization}\label{sec:simu}
The simulation-based solutions for the Influence Maximization problem use the widely-adopted \textit{Monte Carlo} method to estimate the influence function $\sigma(\cdot)$. An MC simulation under the IC model starts from the set $S$, traverses on a \textit{possible world} of $\G$, and returns the number of reachable nodes as an estimation of $\sigma(S)$. To generate a possible world of $\G$, we remove each edge $e=(u,v)$ with probability $1-p(u,v)$. It is a subset sampling problem with $S=\E$, in which each element $e=(u,v)$ is associated by a reserving probability $p(u,v)$. When it comes to dynamic IM problems, edges are inserted into the graph or removed from the graph over time, resulting in element insertions/deletions of the subset sampling problems. 

We conduct experiments on two real-world graphs (OL and TW) with exponential distribution and Weibull distribution and show the query time for possible world generations and the update time. 

For each graph with various skewed distributions, we query for $100$ possible worlds and conduct $1000$ edge insertions and $1000$ edge deletions of the graph. We report the tradeoff between the query time and the update time for all algorithms in Figure~\ref{fig:world}. 
Our \odss outperforms all competitors in terms of both query time and update time across all tested datasets. While the update time of \naive is comparable to our \odss, the query time of \naive is $3\times \sim 5\times$ larger than that of \odss. 
Note that the ratio of the query time of \naive to that of \odss is influenced by the ratio $m/n$, as the query time of \odss in this setting is $O(n)$. Specifically, we scale the sum of probabilities of the outgoing neighbors of each node to $1$, resulting in a sum of probabilities $\mu$ equal to $n$. Therefore, the query time of \odss is $O(1+\mu)=O(n)$, while the query time of \naive is $O(m)$. 
The BringmanSS method performs $2\times$ faster than \naive for a query. However, it needs $100\times$ update time larger than \odss on the OL graph.


\section{Technical Details} \label{sec:proof}


\subsection{Proof of Lemma~\ref{lemma:group}}\label{proof:group}
\begin{proof}
We first consider the correctness of the algorithm. 
Let $R_i$ be the $i$-th random number generated in Algorithm~\ref{alg:group}, that is, $\Pr[R_1=j]={2^{-k+1}(1-2^{-k+1})^{j-1}}/{p(G_k)}$ and $\Pr[R_i=j]=2^{-k+1}(1-2^{-k+1})^{j-1}$ for $1<i\le n_k$. 
Thus, the index of the $i$-th candidate is $\sum_{h=1}^i R_h$.
Note that we call Algorithm~\ref{alg:group} if and only if $G_k$ is sampled with probability $p(G_k)$, i.e., $Y_k=1$. 
Denote the $i$-th candidate as $X_i$. We have 
$$\Pr[X_i=j|Y_k=1]=\Pr\left[\sum_{h=1}^i R_h=j\right].$$
For simplicity's sake, we denote the $j$-th element in $G_k$ as $e_j$. 
Thus, $e_j$ is sampled as the $i$-th candidate with probability 
$$
\Pr[X_i=j]=\Pr\left[\sum_{h=1}^i R_h =j\right] \cdot \Pr[Y_k=1],
$$ for $1\le i\le j.$
Let $Z_i=\sum_{h=1}^{i} R_h$. Note that $e_j$ can be sampled as a $i$-th candidate, where $1\le i\le j$. Then, the probability that $e_j$ is sampled as a candidate is 
$$
\sum_{i=1}^{j} \Pr[X_i=j] = \Pr\left[Y_k=1\right] \sum_{i=1}^{j} \Pr[Z_i=j].
$$
Define $ \hat{p}(e_j) = \sum_{i=1}^j \Pr[Z_i=j] $.
We will prove that $\hat{p}(e_j)=2^{-k+1}/p(G_k)$ for all $j$ using mathematical induction. For the initial step, we have $\hat{p}(e_1)=\Pr[Z_1=1]=2^{-k+1}/p(G_k)$ for the first element $e_1$. Next, for the inductive step, we prove that $\hat{p}(e_j)=2^{-k+1}/p(G_k)$ if the proposition is true for any $1\le j'< j$. 
Consider the probability of $Z_i=j$.
For $i>2$,
$\Pr[Z_i=j]$ can be written as $\sum_{h=i-1}^{j-1}\Pr[Z_{i-1}=h \cap R_i=j-h]$. Note that the random variables $R_1, R_2, \ldots$ are mutually independent. Thus, we have 
\begin{equation}\label{eq:z_i}
\Pr[Z_i=j]=\sum_{h=i-1}^{j-1} \Pr[Z_{i-1}=h]\Pr[R_i=j-h].
\end{equation}
Applying \eqn{\ref{eq:z_i}} to $\hat{p}(e_j)$, we have
\begin{equation*}
    \hat{p}(e_j)=\Pr[Z_1=j]+ \sum_{i=2}^j \sum_{h=i-1}^{j-1} \Pr[Z_{i-1}=h] \cdot 2^{-k+1}(1-2^{-k+1})^{j-h-1}.
\end{equation*}
We interchange the order of the summations and obtain
\begin{equation*}
    \hat{p}(e_j)=\Pr[Z_1=j]+ \sum_{h=1}^{j-1} \sum_{i=1}^h  \Pr[Z_i=h]\cdot 2^{-k+1}(1-2^{-k+1})^{j-h-1}.
\end{equation*}
Note that $\sum_{i=1}^h \Pr[Z_i=h]$ is exactly $\hat{p}(e_h)$. Using the inductive hypothesis, we have $\hat{p}(e_h)=2^{-k+1}/p(G_k)$ for $h<j$. Thus,
\begin{equation}\label{eq: e_j}
    \hat{p}(e_j)=\Pr[Z_1=j]+ \sum_{h=1}^{j-1} \frac{2^{-k+1}}{p(G_k)}\cdot 2^{-k+1}(1-2^{-k+1})^{j-h-1}.
\end{equation}
For $i=1$, we have
\begin{equation}\label{z_1}
\Pr[Z_1=j]=\frac{2^{-k+1}(1-2^{-k+1})^{j-1}}{p(G_k)}.
\end{equation}
Applying \eqn{\ref{z_1}} to \eqn{\ref{eq: e_j}}, we obtain that 
$\hat{p}(e_j)=2^{-k+1}/p(G_k)$. This completes the inductive step. Hence, $e_j$ is sampled with probability
$$
\sum_{i=1}^{j} \Pr[X_i=j]=p(G_k)\cdot 2^{-k+1}/p(G_k)=2^{-k+1}.
$$
We accept each candidate $e_j$ with probability $p(e_j)/2^{-k+1}$. Thus, the element $e_j$ is sampled with probability 
$2^{-k+1}\cdot p(e_j)/2^{-k+1}=p(e_j)$. In a word, given that $G_k$ is sampled with probability $p(G_k)$, Algorithm~\ref{alg:group} returns an unbiased sample of $G_k$.

Next, we consider the expected query time of Algorithm~\ref{alg:group}.
The expected query time only depends on the number of times we generate random numbers, which equals the number of candidates plus one (one time for the index out of the group). 
For $0\le k<K$, the expected number of candidates in $G_k$ is $\sum_{i=1}^{n_k} 2^{-k+1}\le  \sum_{x_i\in G_k} 2p(x_i)=2\mu_k$ since $p(x_i)\le 2^{-k+1}$. For $G_K$, the expected number of candidates is $2^{-K+1}\cdot n_K\le 2^{-K+1}\cdot n\le 1$. Therefore, the expected query time for $G_k$ is $2\mu_k+1=O(\mu_k+1)$ for $0\le k<K$ , and $O(1)$ for $k=K$.
For simplicity, we use $O(\mu_k+1)$ as the expected time for $0\le k\le K$ in the remainder of the manuscript.
\end{proof}

\subsection{Proof of Lemma~\ref{lemma:mu}}
\begin{proof}
$\mu^{(0)}=\mu=\sum_{i=1}^n p(x_i)$. Let $n^{(\ell)}=|S^{(\ell)}|$ be the number of elements at level $\ell$, $n^{(\ell)}_k=|G_k^{(\ell)}|$ be the number of elements contained in $G_k^{(\ell)}$, $K^{(\ell)}=\lceil \log n^{(\ell)}\rceil +1 $ be the largest index of the groups at level $\ell$. 
Note that $p\big(x_k^{(\ell)}\big)=p\big(G^{(\ell-1)}_k\big)$, so $\mu^{(\ell)}=\sum_{k=1}^{K^{(\ell-1)}} p\big(G_k^{(\ell-1)}\big)$.
And $p\big(G_k^{(\ell-1)}\big)=1-(1-2^{-k+1})^{n_k^{(\ell-1)}}$is the probability that $G_k^{(\ell-1)}$ contains at least one candidate. Thus, the sum of the probabilities at level $\ell$ is
$$
\mu^{(\ell)}=\sum_{k=1}^{K^{(\ell-1)}} 1-(1-2^{-k+1})^{n_k^{(\ell-1)}} \le \sum_{k=1}^{K^{(\ell-1)}}  2^{-k+1}n_k^{(\ell-1)}
$$
since $(1-x)^y\ge 1-xy$ for any $0 \le x<1$ and $y\ge 1$. Note that $p\big(x_i^{(\ell)}\big)>2^{-k}$  if $x_i^{(\ell)}\in G_k^{(\ell)}$ for $0\le k<K^{(\ell)}$ at any level $\ell$. Thus, we have 
$$
2^{-k+1}n_k^{(\ell-1)}\le \sum_{x_i\in G_k^{(\ell-1)}} 2p\big(x_i^{(\ell-1)}\big)
$$
for $0\le k<K^{(\ell-1)}$. For $k=K^{(\ell-1)}$, $2^{-k+1}=2^{-\lceil \log n^{(\ell-1)}\rceil}\le 1/n^{(\ell-1)} $. Note that $n_k^{(\ell-1)} \le n^{(\ell-1)}$, so $2^{-k+1}n_k^{(\ell-1)}\le 1$ for $k=K^{(\ell-1)}$. Therefore, we have 
$$
\mu^{(\ell)} \le\sum_{x_i^{(\ell-1)}\in S^{(\ell-1)}} 2p(x_i^{(\ell-1)})+1 = 2\mu^{(\ell-1)}+1.
$$
Hence, we can derive the upper bound of $\mu^{(\ell)}$: 
\begin{equation*}
\mu^{(\ell)}\le2\mu^{(\ell-1)}+1\le 2(2\mu^{(\ell-2)}+1)+1 \le \cdots\le 2^\ell\mu+2^\ell-1.
\end{equation*} 
\end{proof}

\subsection{Proof of Theorem~\ref{theorem:bdss}}
\begin{proof}
We first show that Algorithm~\ref{alg:bdss} draws an unbiased sample of $S$. Firstly, the Naive method gives an unbiased sample of $S^{(L)}$. Each sampled element $x_k^{(L)}$ indicates that $G^{(L-1)}_k$ is sampled with $p\big(G^{(L-1)}_k\big)$. According to Lemma~\ref{lemma:group}, given that a group $G_k^{(\ell)}$ is sampled with $p\big(G_k^{\ell}\big)$, Algorithm~\ref{alg:group} returns a sample of $G_k^{(\ell)}$. Thus, we obtain an unbiased sample of $S^{(L-1)}$ by combined the outcomes of {\sf SamplingWithinGroup}($G^{(L-1)}_k$) for all sampled groups $G^{(L-1)}_k$ . Repeating in this fashion, we finally obtain an unbiased sample of $S^{(0)}$. Thus, Algorithm~\ref{alg:bdss} returns an unbiased sample of the subset sampling problem.

Next, we prove the expected query time of Algorithm~\ref{alg:bdss}.
Note that $L$ equals $\log^*n$ to guarantee the number of elements at level $L$ is a constant. Sampling elements at level $L$ (Algorithm~\ref{alg:bdss} Line~\ref{line:naive}) costs constant time since $n^{(L)}$ is a constant. 
Consider the cost of sampling elements at level $\ell$ for $\ell<L$, denoting as $C^{(\ell)}$, given that some groups at level $\ell$ are sampled (Algorithm~\ref{alg:bdss} Line~\ref{line:while}~to~\ref{line:group}). Note that the expected number of sampled groups at level $\ell$ is exactly the expected number of elements sampled at level $\ell+1$, which equals $\mu^{(\ell+1)}$. According to Lemma~\ref{lemma:group}, Algorithm~\ref{alg:group} {\sf SamplingWithinGroup}\big($G^{(\ell)}_k$\big) costs $2\mu^{(\ell)}_k+1$ expected time for $0\le k\le K^{(\ell)}$, where $\mu^{(\ell)}_k$ is the sum of probabilities for the elements in $G^{(\ell)}_k$. Let $T$ be the set of the indexes of the sampled groups. We have
$$
\Ex\big[C^{(\ell)}\big]\le O(1)+\mu^{(\ell+1)}+\Ex\left[\sum_{k\in T} (2\mu^{(\ell)}_k+1)\right],
$$
in which the $O(1)$ term for miscellaneous overhead, the $\mu^{(\ell+1)}$ term for iterating in the $\mu^{(\ell+1)}$ sampled groups. 
According to the linearity of expectation, we have $\Ex\big[\sum_{k\in T} (2\mu_k^{(\ell)}+1)\big]=\Ex\big[\sum_{k\in T}2\mu^{(\ell)}_k\big]+\Ex[\sum_{k\in T} 1]$. 
The first term $\Ex\big[\sum_{k\in T} 2\mu^{(\ell)}_k\big]$ is less than the sum of $2\mu_k$ for all $0\le k\le K^{(\ell)}$, which equals $2\mu^{(\ell)}$. The second term $\Ex[\sum_{k\in T} 1]$ equals $\mu^{(\ell+1)}$. Thus, $\Ex[C^{(\ell)}]\le 2\mu^{(\ell+1)}+2\mu^{(\ell)} +O(1)$. The total query cost can be bounded by
$$\Ex[C]=\Ex\left[\sum_{\ell=0}^L C^{(\ell)}\right]\le O(1)+\sum_{\ell=0}^{L-1} \left(2\mu^{(\ell+1)}+2\mu^{(\ell)}+O(1)\right). $$
Applying \eqn{\ref{eq:mu}}, we derive the expected query time
$$
\Ex[C]=O\left(2^{\log^*n}\mu+2^{\log^*n}+\log^*n\right).
$$

Finally, we prove the preprocessing time and the memory cost of Algorithm~\ref{alg:bdss}. In the preprocessing phase, we check the probability of each element and then assign it to an appropriate group at each level $\ell$ (except for $L$ as $G^{(L)}$ is not necessary). 
Then, the probabilities of the groups are computed to obtain the probabilities of elements at the next level. Note that $p\big(G_k^{(\ell)}\big)=1-(1-2^{-k+1})^{n_k^{(\ell)}}$ can be calculated in constant time since $a^b=\exp(b\log a)$. 
And the number of the groups at level $\ell$ is $\lceil \log n^{(\ell)} \rceil+1$, which equals $n^{(\ell+1)}$. 
Thus, the cost of each level $\ell$ is $O(n^{(\ell)}+n^{(\ell+1)})= O(n)$. 
The total cost of preprocessing is $\sum_{\ell=0}^{L-1} \big(n^{(\ell)}+n^{(\ell)}\big)= O(nL)=O(n\log ^*n)$. 
Similary, the cost of memory space at level $\ell$ is $O\big(n^{(\ell)}+\lceil \log n^{(\ell)} \rceil+1\big)$ for $n^{(\ell)}$ elements and $\lceil \log n^{(\ell)} \rceil+1$ groups. 
The total memory space is $O(n\log ^*n)$.
\end{proof}

\subsection{Proof of Lemma~\ref{lemma:table}}
\begin{proof}
We first show that the table lookup method gives an unbiased sample of $S$. 
Denote the subset as $B$ that we obtain by indexing into the table with $A ( \bar{p}(x_1), \ldots, \bar{p}(x_{m}) ) $ and a uniformly distributed random $r$. 
$B$ is selected with $p(B)$ as we fill in $m^mp(B)$ entries with $B$ in the total $m^m$ entries of the row. 
For each element $x_i$, it is included in $B$ with probability $\bar{p}(x_i)$. If $x_i$ is included in the selected $B$, that is, $x_i$ is sampled as a candidate, we accept it with $p(x_i)/\bar{p}(x_i)$. Thus, $x_i$ is sampled with $\bar{p}(x_i)\cdot p(x_i)/\bar{p}(x_i)=p(x_i)$. The unbiasedness of the table lookup method follows.

Next, we consider the expected query time of the table lookup method. When querying, we generate a random $r$ uniformly from $\{ 0, \ldots, m^m - 1\}$ and obtain the bit array stored in the entry in row $A(\bar{p}(x_1),\ldots,\bar{p}(x_{m}))$ and column $r$.
To decode the bit array, we first get the position $i$ of the rightmost $1$-bit by standard bit operations in $O(1)$ time.
$x_{i}$ is the sampled candidate with the smallest index. 
Then, replace $B$ with $B-2^i$. 
Repeating the technique above with $B$, we will get all the sampled candidates. 
For each candidate, we then use rejection to get the probability of candidate $x_i$ down to $p(x_i)$. 
Thus, the query time depends on the number of sampled candidates. 
Note that $\bar{p}(x_i)-p(x_i)\le\frac{1}{m}$ since $\bar{p}(x_i)=\lceil m p(x_i)\rceil/m$. Thus we have 
$$
\sum_{i=1}^m \bar{p}(x_i)\le \sum_{i=1}^m \frac{1}{m}+p(x_i)\le \mu+1,
$$
where $\mu=\sum_{i=1}^{m} p(x_i)$. Thus, the expected query time is $O(\mu+1)$.

At the preprocessing phase, we maintain the table and calculate $A(\bar{p}(x_1),\ldots,\bar{p}(x_{m}))$ for the current $(p(x_1),\ldots,p(x_{m}))$. For each table row, we calculate the probabilities of $2^m$ subsets. Note that each probability of a subset is the product of $m$ probabilities, which costs $O(m)$ time to calculate. The time for filling in the table is $O(m^m\cdot m^m)$. Calculating $A(\bar{p}(x_1),\ldots,\bar{p}(x_{m}))$ costs $O(m)$ time. Thus, the total preprocessing time is $O(2^m\cdot m+m^{2m})$.

The memory space depends on the size of the table, which is $O(m^{2m})$. Note that the word length is at least $\log m^m$ bits to store the row index in a single memory word. We will show in the proof of Theorem~\ref{theorem:dss} that the word length under the standard word RAM model is sufficient since the number of elements $m$ is so few.
\end{proof}

\subsection{Proof of Theorem~\ref{theorem:dss}} \label{proof:dss}
\begin{proof}

We first show the correctness of Algorithm~\ref{alg:dss}. This optimal algorithm is derived by replacing the Naive method in Algorithm~\ref{alg:bdss} with the table lookup method. Since we have proved the correctness of the table lookup method in Lemma~\ref{lemma:table} and the correctness of the basic algorithm in Theorem~\ref{theorem:bdss}, the correctness of Algorithm~\ref{alg:dss} follows.

Consider the expected query time. As proved in Theorem~\ref{theorem:bdss}, the expected cost of sampling elements within the sampled groups at level $\ell$ is $\Ex[C^{(\ell)}]\le 2\mu^{(\ell+1)}+2\mu^{(\ell)}+O(1)$. According to Lemma~\ref{lemma:table}, the expected cost of table lookup at level $2$ is $\Ex[C^{(2)}]=1+\mu^{(2)}$. Thus, the expected query time of Algorithm~\ref{alg:dss} is 
$$
\Ex[C]=\Ex[C^{(0)}]+\Ex[C^{(1)}]+\Ex[C^{(2)}]\le 3\mu^{(2)}+4\mu^{(1)}+2\mu+3.
$$
According to Lemma~\ref{lemma:mu}, $\mu^{(\ell)}\le 2^{\ell}\mu+2^{\ell}-1$. Thus, we derive the expected query time of Algorithm~\ref{alg:dss} as 
$$
\Ex[C]=O(1+\mu). 
$$

Consider the expected update time. As we illustrate in Appendix~\ref{sec:app_structure}, an insertion/deletion of an element within a group can be done in constant time. Each update operation (element insertion/deletion, modification of probability) can be resolved into several insertions or deletions within a group. Thus, all update operations can be done in constant time. 

The preprocessing time for level $1$ and level $0$ is $O(n)$ as proved in Theorem~\ref{alg:bdss}. To support efficient updates, we maintain an array for storing the position in the corresponding group for each element $x_i$. The total preprocessing time remains $O(n)$ for level $1$ and level $0$. According to Lemma~\ref{lemma:table}, the preprocessing time for the table lookup method for level $2$ is $O(2^m\cdot m+m^m)$ when there are $m$ elements at level $2$. Note that $m=\lceil \log (\lceil \log n \rceil +1)\rceil +1$, and hence $O(2^m\cdot m+m^m)\le O(n)$. Thus, the preprocessing time for each level $0\le\ell\le 2$ is $O(n)$, so is the total preprocessing time. 

At level $0$ and level $1$, we maintain an array for each group to store the containing elements. To help with locating the position of the elements in the groups, we also maintain another array at level $0$ and level $1$ to store the position of each element. Thus, the total memory space actually in use is $O(n)$. 
According to Lemma~\ref{lemma:table}, it costs $O(m^{2m})$ space for level $2$. Since $m=\lceil \log (\lceil \log n \rceil +1)\rceil +1$, we have $O(m^{2m})\le O(n)$. As mentioned in the proof of Lemma~\ref{lemma:table}, the table lookup method requires the word length of at least $\log m^m$ bits. Since $m=O(\log\log n)$, the word length of $O(\log n)$ bits in the word RAM model is sufficient. Therefore, we conclude that the total memory space for the optimal algorithm \odss is $O(n)$.
\end{proof}

\end{document}